\documentclass[]{elsarticle}

\usepackage[official]{eurosym}
\usepackage{todonotes}
\usepackage{comment}
\usepackage{graphicx}
\usepackage{makecell}
\usepackage{tabularx}
\usepackage{longtable}
\usepackage{booktabs,cellspace}
\usepackage{enumitem}
\usepackage[hyphens]{url}
\usepackage{hyperref}
\usepackage{multirow}

\usepackage{amsthm}

\makeatletter
\def\thmhead@plain#1#2#3{%
  \thmname{#1}\thmnumber{\@ifnotempty{#1}{ }\@upn{#2}}%
  \thmnote{ {\the\thm@notefont#3}}}
\let\thmhead\thmhead@plain
\makeatother
\theoremstyle{break}

\newcommand{\mypar}[1]{\smallskip\noindent\textbf{#1.}}

\newtheorem{example}{Example}
\newtheorem{scenario}{Scenario}

\definecolor{darkgreen}{rgb}{.0,0.53,.0}

\journal{Information Systems}

\makeatletter
\def\namedlabel#1#2{\begingroup
    #2%
    \def\@currentlabel{#2}%
    \phantomsection\label{#1}\endgroup
}
\makeatother

\begin{document}

\begin{frontmatter}

\title{Predictive Compliance Monitoring in Process-Aware Information Systems: State of the Art, Functionalities, Research Directions}

\author{Stefanie Rinderle-Ma, Karolin Winter, Janik-Vasily Benzin}

\address{Technical University of Munich, Germany; TUM School of Computation, Information and Technology, Department of Informatics \\\{stefanie.rinderle-ma, karolin.winter, janik.benzin\}@tum.de}

\begin{abstract}

Business process compliance is a key area of business process management and aims at ensuring that processes obey to compliance constraints such as regulatory constraints or business rules imposed on them. Process compliance can be checked during process design time based on verification of process models and at runtime based on monitoring the compliance states of running process instances. For existing compliance monitoring approaches it remains unclear whether and how compliance violations can be predicted, 
although predictions are crucial in order to prepare and take countermeasures in time. This work, hence, analyzes existing literature from compliance monitoring as well as predictive process monitoring and provides an updated framework of compliance monitoring functionalities. Moreover, it raises the vision of a comprehensive predictive compliance monitoring system that integrates existing predicate prediction approaches with the idea of employing PPM with different prediction goals such as next activity or remaining time for prediction and subsequent mapping of the prediction results onto the given set of compliance constraints (PCM).   
For each compliance monitoring functionality we elicit PCM system requirements and assess their coverage by existing approaches. 
Based on the assessment, open challenges and research directions realizing a comprehensive PCM system are elaborated. 

\end{abstract}


\end{frontmatter}

\section{Introduction}\label{sec:introduction}

The need for predictive and online data analysis is crucial given the highly volatile economic environment in which processes 
have to constantly adapt to new circumstances, e.g., to COVID-19 circumstances or the Ucraine war and, thus, historical data may be outdated and proactive process management gains importance \cite{Pane2021,DBLP:conf/ecis/StierleBWZM021,breuker_comprehensible_2016,poll_process_2018}. 
In the area of business process management, Predictive Process Monitoring (PPM) \cite{DBLP:conf/bpm/Francescomarino18,DBLP:journals/tist/VerenichDRMT19,DBLP:journals/tsc/Marquez-Chamorro18} has attained tremendous interest recently and several approaches for predicting, for example, the remaining time of cases, the next activity, or the outcome of a process have been presented. Doing so, PPM can be a valuable means for estimating company-relevant key performance indicators, for example, customer retention. 
Hence, PPM can enable the proactive management of business processes and operational risks \cite{conforti_prism_2016}. 

In addition to PPM, Compliance Monitoring (CM) \cite{DBLP:journals/is/LyMMRA15,tambotoh_process_2021} is an integral part for monitoring and managing business processes in changing, complex regulatory environments such as the financial domain. By combining the respective capabilities of PPM and CM, research can offer companies a means to proactively assess and manage their business processes with respect to future outcomes, compliance status, and risks. Yet, reactive management through auditing is still most prominent in compliance management of companies, explaining why the use of predictive data analysis is an important factor for companies to improve 
and underlines the need in the current environment to bring both lines of research, i.e., PPM and CM, together \cite{amy_matsuo_kpmg_2021}.

So far, the combination of PPM and CM has not been put to the test, i.e., it has not been systematically analyzed what capabilities of CM are already covered by PPM and what may be missing to fully support 
\textsl{Predictive Compliance Monitoring} (PCM), i.e. predictive compliance management and monitoring in an online setting \cite{DBLP:journals/is/LyMMRA15} where
compliance violations of process instances are predicted during runtime.
For CM, a framework of functionalities called Compliance Monitoring Functionalities (CMF) \cite{DBLP:journals/is/LyMMRA15} has been established that serves as means to systematically compare and analyze existing approaches. 
Consequently, the CMFs are well-suited as a starting point for testing the combination of PPM and CM, ideally supported by a comprehensive \textsl{PCM system}.

\begin{figure}
    \centering
    \includegraphics[width=\textwidth]{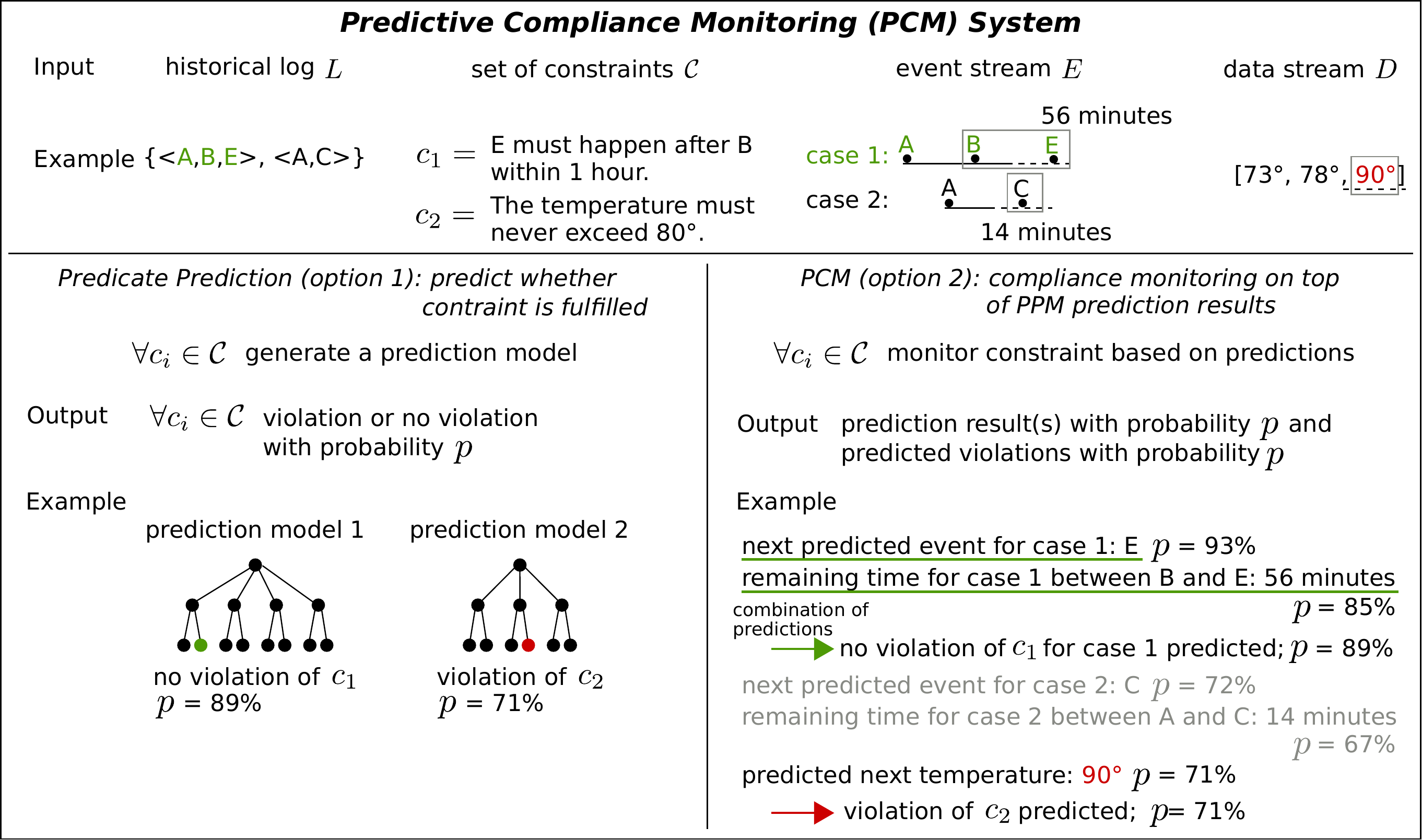}
    \caption{Predictive Compliance Monitoring System -- Problem Statement and Example}
    \label{fig:overview_pcm_system}
\end{figure}

Building a PCM system is a complex task, resulting from a multitude of compliance constraints stemming from different and constantly changing regulatory documents \cite{DBLP:journals/kais/HashmiGLW18} and referring to multiple process perspectives \cite{DBLP:journals/tsc/CabanillasRR22}, a process event log/stream that is emitted from multiple, heterogeneous sources/systems \cite{DBLP:conf/smds/Aalst21}, as well as regular and irregular changes of the process \cite{DBLP:conf/IEEEscc/MaisenbacherW17}. Moreover, it should be possible to integrate an existing PPM approach seamlessly into a PCM system as depicted in Figure \ref{fig:overview_pcm_system}. \textsl{Predicate prediction} -- which is mainly followed by existing approaches, e.g., \cite{DBLP:conf/caise/MaggiFDG14} -- encodes each compliance constraint as prediction goal into prediction models. Often the predicates are encoded using formalisms such as Linear Temporal Logic (LTL), but basically any function that computes a target value based on a trace in the log or stream is sufficient, if it can be related one-to-one with a constraint that expresses what the target value means with respect to the compliance state of the trace. The other option, i.e., \textsl{PCM}, integrates PPM approaches with possibly different prediction goals such as next activity, remaining time, outcome, or any indicators as well as combinations of them and then maps the prediction results onto the set of constraints $\cal{C}$. An example for a constraint requiring combination of next activity and temporal prediction is constraint $c_1$. An example for a constraint that requires prediction of data values, in this case the temperature based on an external data stream, is $c_2$. By contrast to CM, predicate prediction and PCM augment compliance violation predictions with probabilities. Augmenting compliance violation predictions with probabilities entails a decisive change with respect to decision making by the company, as these probabilities lead to decision making under risk or uncertainty \cite{mousavi2014risk}. 
 
Ideally, a PCM system is able to support both, predicate prediction and PCM, and it is crucial to understand for which cases of which option is preferable, e.g., regarding prediction quality. This requires PCM to be a continuous task and a PCM system to support continuous compliance prediction and re-evaluation of previous results based on different criteria such as constraint priorities which could depend, e.g., on the expected fine upon violation.

The complexity of a PCM system can be illustrated based on an example from the financial domain, i.e., the transaction reporting processes of financial institutions in the EU\footnote{A detailed description of the example can be found in the appendix, cf. Example \ref{ex:eu}.} \cite{thomas_wenzel_transaction_2020}. These processes are subject to a multitude of regulations such as the European Markets Infrastructure Regulation \cite{european_parliament_regulation_2014}, Money Markets Statistical Reporting \cite{european_parliament_regulation_2015}, Securities Financing Transactions Regulation \cite{european_parliament_regulation_2019}, and Markets in Financial Instruments (MiFIR/D) \cite{european_parliament_regulation_2018} which are complemented by addenda and technical specification documents, e.g., reporting instructions \cite{european_central_bank_reporting_2021}, questions and answers \cite{european_central_bank_money_2021}, IT appendix \cite{european_central_bank_mmsr_2017}, data quality checks \cite{european_central_bank_moneymarket_2021} and further technical specifications documents. Together with event streams from different sources, a PCM system has to integrate all information in an efficient and usable fashion. 

This work addresses the question to which extent a PCM system in its entirety is addressed and solved by existing PPM and CM approaches (\textbf{RQ1}), how the existing PPM and CM approaches are comparable in terms of CMF functionalities, in particular, PCM system requirements necessary for the respective functionality (\textbf{RQ2}), which of the requirements imposed by a PCM system are already met by existing approaches (\textbf{RQ3}), and which challenges and research directions remain still open (\textbf{RQ4}). 

The remainder of the paper is structured as follows.
We start with a compilation and review of PPM, CM, and PCM literature ($\mapsto$ RQ1). Afterwards, the compiled literature is analyzed along the existing framework on CMFs \cite{DBLP:journals/is/LyMMRA15} and for possible extensions to this initial framework ($\mapsto$ RQ2). Analogously to \cite{DBLP:journals/is/LyMMRA15} literature is complemented by case studies as source for the CMF extension. Finally, we compare the findings from the literature study with two recent compliance surveys \cite{DBLP:journals/kais/HashmiGLW18,DBLP:journals/tsc/CabanillasRR22}. Based on the analysis of PPM, CM and PCM literature, we investigate the relation of PPM and PCM, i.e., whether i) PPM can fully encompass PCM by defining compliance violations as prediction goal (predicate prediction) or ii) PPM is utilized for predicting compliance violations based on predicting, e.g., next activities or remaining time, and interpreting the prediction results, even in combination, over the set of compliance constraints (PCM). The analysis of the literature compilation combined with findings from case studies results in an extended CMF framework. The extended CMF framework is then analyzed for the PCM system requirements arising for each of the CMFs, illustrated by means of
Example \ref{ex:eu} ($\mapsto$ RQ2). 
The CMFs, together with their PCM system requirements, are then analyzed for support by existing (mostly PPM) approaches, categorized along their prediction goals, e.g., next activity or outcome ($\mapsto$ RQ3). Then, we discuss the implications of both the extended CMF framework, its PCM system requirements and the assessment of existing approaches for the relationship of PPM, CM and PCM as a combination of the former two ($\mapsto$ RQ3). Furthermore, we provide suggestions how each CMF can be addressed to elucidate the relationship of the different research areas. Lastly, we provide a set of open challenges and research directions ($\mapsto$ RQ4) that emerge from the assessment and discussion.

The contributions to tackle RQ1 -- RQ4 comprise literature reviews for PCM, CM, and PPM, resulting in an extended CMF framework  (cf. Sect. \ref{sec:literature_compilation}). The extended CMF framework is described and illustrated in Sect. \ref{sec:prediction_requirements} and serves as basis for deriving requirements for developing a comprehensive PCM solution. Furthermore, existing PPM approaches are put to the test along these requirements. 
The analysis steps finally culminate in open challenges and research directions for PCM (cf. Sect. \ref{sec:research_agenda}). Limitations of this survey are discussed in Sect. \ref{sec:discussion} and results concluded in Sect. \ref{sec:conclusion}. 

The extended CMF framework and the research directions provide several open research topics from a data, algorithmic, and application perspective for PPM and PCM. Overall, this work aims at bridging the gap between online and predictive process analysis techniques and real-world compliance management.

\section{Literature Review}\label{sec:literature_compilation}

\begin{figure}
    \centering
    \includegraphics[width=\textwidth]{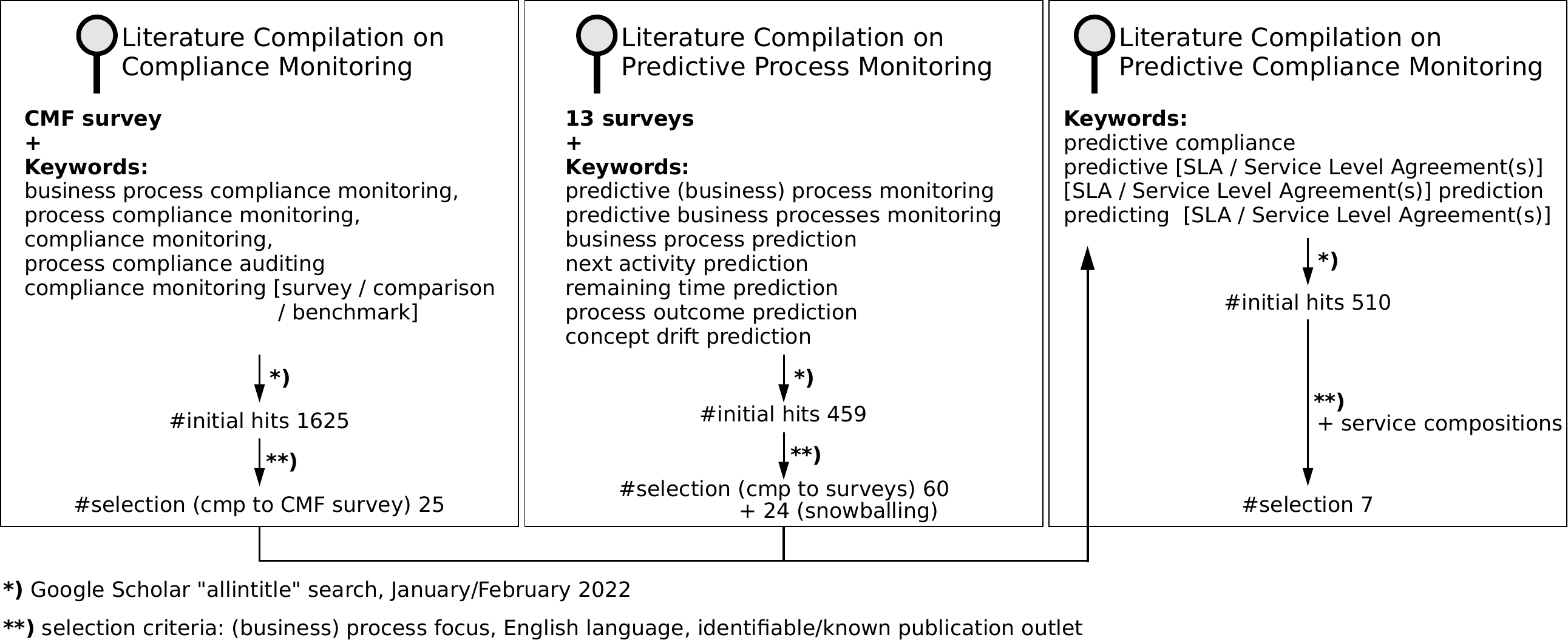}
    \caption{Search Methodologies for the Literature Compilations on CM, PPM and PCM}
    \label{fig:lit_comp_wf}
\end{figure}

Figure \ref{fig:lit_comp_wf} depicts the literature search and selection methodology applied in this paper.  
We start with literature on compliance monitoring (CM), followed by literature on predictive process monitoring (PPM), complemented by a search for predictive compliance monitoring (PCM) approaches.
The literature lists are available via \url{https://www.cs.cit.tum.de/bpm/data/}.

The overall goal of the literature review is to assess whether and to which extent PCM is addressed by existing approaches and to set out a research agenda for PCM and in particular a PCM system. This necessitates building a basis for the assessment, i.e., a set of PCM system requirements based on which existing approaches can be evaluated and potential research gaps can be identified. We use the well-established Compliance Monitoring Functionality framework presented in \cite{DBLP:journals/is/LyMMRA15} as basis and update and extend it with respect to predictive requirements. 
The original framework \cite{DBLP:journals/is/LyMMRA15} defines the following Compliance Monitoring Functionalities (CMFs):
\begin{itemize}
    \item \textsl{Modeling requirements:} CMF1 (time), CMF2 (data), CMF3 (resources)\\ CMF1--3 refer to the modeling capabilities of the compliance constraints. The underlying assumption is that all compliance constraints refer to the control flow of a process, e.g., by referring to the existence of an activity plus a maximal duration of this activity ($\mapsto$ CMF1). 
    \item \textsl{Execution requirements:} CMF4 (non-atomic activities), CMF5 (life cycles), CMF6 (multiple instances constraints)\\
    CMF4--6 refer to instantiation and execution of the process instances, more precisely the event and life cycle information that is stored in the process event streams during runtime, and the instantiation of the compliance constraints. 
    \item \textsl{User requirements:} CMF7 (reactive management), CMF8 (proactive management), CMF9 (explain root cause of violation), CMF10 (quantify compliance degree)\\
    CMF7--10 refer to support that approaches offer for users to understand and handle compliance violations. CMF8 refers to PCM as proactive management of compliance violations that requires the prediction of such violations. 
\end{itemize}

In the following, we analyze the papers from each literature compilation regarding two aspects, and aggregate the results at the end of this section. 
\begin{itemize}
\item[i)] Which existing CMFs from \cite{DBLP:journals/is/LyMMRA15} are mentioned/addressed?
\item [ii)] Which possible CMF extensions are mentioned/addressed?
\end{itemize}

\subsection{CM Literature Compilation and Findings}\label{subsec:slr_cm}
We take the CM functionality framework and systematic literature survey from $2015$  \cite{DBLP:journals/is/LyMMRA15} as a yardstick, i.e., we assume that CM literature up to $2015$ has been mostly covered by \cite{DBLP:journals/is/LyMMRA15}. For keywords and selection criteria see Fig. \ref{fig:lit_comp_wf}.
The search resulted in $1625$ initial hits. From these $1625$ papers, $25$ papers were selected as in scope of PCM and not yet analyzed by the CM survey in \cite{DBLP:journals/is/LyMMRA15}. 
The analysis of these $25$ papers results in $17$ papers that mention or address CMFs in the following way: 

\textsl{i) Which CMFs from \cite{DBLP:journals/is/LyMMRA15} are mentioned/addressed by CM approaches?} Most of the existing CM approaches \cite{awad_compliance_2015,barnawi_runtime_2015,barnawi_anti-pattern-based_2016,sakr_compliance_2015,gong_bpcmon_2017,hedjeres_temporal_2018,knuplesch_framework_2017,koenig_compliance_2019,loreti_distributed_2018,maggi_compliance_2019,meroni_multi-party_2018,van_der_werf_online_2015,zaki_enabling_2017,diaz_contract_2013} address modeling requirements CMF1--CMF3 fully or partly through their support for the process perspectives time, data, and resources and depending on the employed constraint modeling formalism such as linear temporal logic (LTL), event-based compliance language (ECL), compliance rule graphs (CRG), and variants of event condition action rules (ECA rules), i.e., timed ECA rules or match condition action rules. Fewer approaches mention or address execution requirements CMF4--CMF6, mostly those approaches that support some kind of activity life cycle \cite{barnawi_runtime_2015,barnawi_anti-pattern-based_2016,sakr_compliance_2015,knuplesch_framework_2017,koenig_compliance_2019,meroni_multi-party_2018,zaki_enabling_2017}. Several approaches address user requirements CMF8--CMF10 \cite{awad_compliance_2015,barnawi_runtime_2015,barnawi_anti-pattern-based_2016,sakr_compliance_2015,knuplesch_framework_2017,knuplesch_framework_2017,maggi_compliance_2019,meroni_multi-party_2018,diaz_contract_2013} based on providing reactive and partly proactive management of compliance violations as well as visualization approaches for compliance states, e.g., satisfied or violated. Overall, it seems that the majority of approaches are basic in the sense that they focus on providing support for compliance monitoring in terms of modeling constraints and checking them over event streams, hence addressing modeling requirements CMF1--CMF3. There are also approaches that aim at providing a comprehensive CM solution/framework by addressing modeling and user requirements. For the execution requirements, the ``coverage'' depends on how approaches are able to deal with different activity life cycles and constraints that span across multiple instances or processes. The latter (CMF6) has not been covered by existing CM approaches yet.

\textsl{ii) Which possible CMF extensions are mentioned by CM approaches?}  \cite{awad_compliance_2015,loreti_distributed_2018} emphasize the efficiency/performance of the CM approach in order to deal with a large volume of events as well as the aspect of data quality.  \cite{comuzzi_alignment_2017,knuplesch_framework_2017,loreti_distributed_2018,meroni_multi-party_2018} raise the requirement to support CM in process collaborations, for example, compliance in connection with the dynamic replacement of partners leaving a process collaboration or new partners joining. The support of CM in distributed settings is also emphasized by \cite{koenig_compliance_2019} in supporting the (semantic) aggregation of events from heterogeneous sources. \cite{gong_bpcmon_2017} advocate the aggregation of values of multiple events and event correlation for addressing multiple data sources. Moreover, the efficiency of the approaches is put into the spotlight for dealing with a large volume of events. 
\cite{maggi_compliance_2019} mentions the \textsl{``early detection of conflicting constraints''} where only one of the constraints can be fulfilled at a time. \cite{awad_compliance_2015,barnawi_runtime_2015,barnawi_anti-pattern-based_2016} emphasize the consistency of the compliance constraint base.

From the analyzed approaches, \cite{kotamarthi_framework_2015, svatos_requirements_2017} as well as the survey in \cite{tambotoh_process_2021} cannot be fully assessed due to lack of technical detail. 

\mypar{Conclusion} CMF1--CMF10 as proposed in \cite{DBLP:journals/is/LyMMRA15} are still valid and approaches since its publication in 2015 address several of the outlined CMFs. After 2015, new directions/requirements include:
    \begin{itemize}
        \item Efficiency/performance of compliance monitoring
        \item Compliance monitoring in distributed processes
        \item Integration of event streams from multiple data sources
        \item Consistency of the constraint base
    \end{itemize}
Efficiency and performance of compliance monitoring is motivated in existing literature by the volume of the event data, also in combination with the applied (ML) technique. The assessment of an approach regarding its efficiency and performance is depending on the application and users, e.g., an CM approach taking $5$ hours can still be efficient if users expect the results within $12$ hours. Hence, the requirement of efficiency/performance will be considered a user requirement. The other requirements refer to \textsl{data}. Hence, CMF1--CMF10 will be extended with one user requirement and new \textsl{data requirements} accordingly. We will describe and illustrate these extensions in Sect. \ref{subsec:extended_cmf}.

\subsection{PPM Literature Compilation and Findings}\label{subsec:slr_ppm}

 For keywords and selection criteria used in the literature search for PPM see Fig. \ref{fig:lit_comp_wf}.
 The search results comprise $459$ initial hits among which $126$ are selected based on the outlined criteria, including $12$ survey papers. The publication dates of the surveys include $2$ surveys in $2018$ \cite{DBLP:conf/bpm/Francescomarino18,DBLP:journals/tsc/Marquez-Chamorro18}, $3$ surveys in $2019$ \cite{DBLP:journals/tist/VerenichDRMT19,teinemaa_outcome-oriented_2019,tama_empirical_2019}, $3$ surveys in $2020$ \cite{spree_predictive_2020,harane_comprehensive_2020,ogunbiyi_comparative_2020}, and $4$ surveys in $2021$ \cite{wolf_framework_2021,DBLP:conf/quatic/KappelJS21,DBLP:conf/ecis/StierleBWZM021,rama-maneiro_deep_2021}. We add the survey in \cite{DBLP:journals/air/NeuLF22} due to snowballing. 
 
 The $13$ surveys provide classifications for PPM approaches based on prediction goals \cite{DBLP:conf/bpm/Francescomarino18,DBLP:journals/tsc/Marquez-Chamorro18,wolf_framework_2021}, techniques \cite{harane_comprehensive_2020, neu_systematic_2021,rama-maneiro_deep_2021}, goals and technique \cite{ogunbiyi_comparative_2020}, and use cases \cite{spree_predictive_2020} as well as benchmarks regarding specific techniques \cite{DBLP:journals/eswa/TamaC19}, a specific prediction goal \cite{DBLP:journals/tkdd/TeinemaaDRM19,DBLP:journals/tist/VerenichDRMT19}, or specific data requirements \cite{spree_predictive_2020}. \cite{DBLP:conf/ecis/StierleBWZM021} provides a classification based on the explainability of PPM approaches. The classifications provided by existing surveys are partly utilized for the analysis conducted in Sect. \ref{sec:prediction_requirements} and comprise prediction goals relevant in the context of PCM, e.g., LTL rules, and for CMF extensions, e.g., inter-case metrics. Also small data sets and explainability point to possibly relevant CMF extensions. 
 
 At first, the analysis of the $13$ surveys results in $10$ papers (partly contained in the $126$ papers in the literature selection) that claim to provide a solution to PCM, i.e., compliance monitoring through PPM. Out of these $10$ approaches, $4$ do not address any constraint definition, $3$ address constraint definition through SLA \cite{castellanos_comprehensive_2004,leitner_data-driven_2013,DBLP:journals/dss/ConfortiLRAH15}, and $3$ address constraint definition in the form of predicates \cite{DBLP:conf/caise/MaggiFDG14,francescomarino_clustering-based_2019,santoso_specification-driven_2020}, e.g., based on LTL constraints. The latter three approaches, in particular, define predicates as prediction goals, i.e., it becomes directly possible to predict violations. \cite{DBLP:journals/tsmc/MetzgerLISFCDP15} outlines three basic PPM approaches based on machine learning, constraint satisfaction, and QoS aggregation which are all relevant for PCM.
We consider SLA to be in scope for the subsequent analysis on PPM and the search on PCM, as the majority of relevant PPM papers targets SLAs. In this context, we also consider \textsl{prescriptive process monitoring} approaches, which are concerned with finding the optimal execution time of interventions to optimize relevant KPIs or meet SLAs of the process such as the cycle time of process instances through predictions at runtime \cite{DBLP:conf/icpm/BozorgiTDRP21,metzger_triggering_2020}. These approaches can be also interesting for possible actions after violations have been predicted. 
  
 Similar to the CM literature review, we use the $13$ surveys as yardstick to distinguish ``non-survey'' papers into
 papers that have already been analyzed and contribute to the conclusions of one ore several of the $13$ surveys, and into papers that have not been analyzed by a survey yet. The latter class of papers comprises $60$ papers. Snowballing yields $24$ PPM papers to be potentially relevant for PCM. 
From these $60+24$ papers, $14+9$ mention/address ``compliance'' or ``SLA'' 
 and go into the following discussion of questions i) and ii) set out in the previous section for CM.

\textsl{i) Which CMFs from \cite{DBLP:journals/is/LyMMRA15} are mentioned/addressed by PPM approaches?} Existing PPM approaches \cite{leitner_data-driven_2013,DBLP:journals/is/LeoniAD16,DBLP:journals/sosym/HeinrichMHP17,DBLP:journals/is/AalstSS11,DBLP:conf/bis/PegoraroUGA21,DBLP:conf/icsoc/WangC21} address modeling requirements CMF1--CMF3 fully or partly as PPM perspectives time, data, and resource. Control flow, by contrast to \cite{DBLP:journals/is/LyMMRA15}, is explicitly mentioned as PPM perspective, especially in the context of next activity/event prediction approaches such as \cite{DBLP:conf/ijcnn/CuzzocreaFGP16,DBLP:journals/dss/EvermannRF17,DBLP:journals/tsc/Pasquadibisceglie22} as well as pattern \cite{DBLP:conf/iceis/BevacquaCFGP13} and predicate prediction approaches \cite{francescomarino_clustering-based_2019}. 
The support of execution requirements CMF4--CMF5 is missing as existing approaches have not supported any kind of (activity) life cycle yet. CMF6 is addressed by \cite{DBLP:conf/bpm/SenderovichFGJM17} through predictions considering intra-case and inter-case features. User requirement CMF8 is addressed  \cite{DBLP:journals/corr/abs-2109-03501,leitner_data-driven_2013} by, e.g., providing BPI cockpits \cite{castellanos_comprehensive_2004} or suggesting mitigation actions \cite{dumas_enabling_2015}. 
Several PPM approaches focus on the explainability of the prediction results \cite{DBLP:conf/otm/BohmerR18,mehdiyev_novel_2020,DBLP:conf/ispw/VerenichNRD17}. They can be mapped onto CMF9 on explaining root causes for compliance violations as proposed in \cite{DBLP:journals/is/LyMMRA15}, but CMF9 can be refined into more precise CMFs, i.e., 1) root cause analysis and 2) effective communication of root cause as in \cite{DBLP:journals/is/LyMMRA15}, as well as additionally in 3) explaining and visualizing prediction results, 4) explaining and visualizing the set of future violations, and 5) explaining and visualizing the effects of mitigation actions on predicted/future violations. Moreover, we advocate to rename CMF9 into \textsl{CMF9': Explainability}.  

\textsl{ii) Which possible CMF extensions are mentioned by PPM approaches?} \cite{DBLP:journals/dss/ConfortiLRAH15,DBLP:conf/wecwis/Brunk20,DBLP:journals/is/BorkowskiFNRS19,DBLP:conf/otm/FolinoGP12}
mention the requirement to consider external (process) context data. This can be underpinned by other recent approaches such as \cite{DBLP:conf/bpm/StertzRM20,DBLP:conf/icsoc/EhrendorferMR21} showing that context can provide useful information for root cause analysis and explainability in PPM. \cite{DBLP:conf/IEEEscc/MaisenbacherW17} mentions \textsl{concept drift} as crucial factor to be considered in PPM. Prediction in distributed processes is addressed in \cite{DBLP:journals/is/BorkowskiFNRS19}. 
Finally, PPM approaches address the properties and quality of the data,
i.e., the event streams, including the size of the input data \cite{DBLP:journals/infsof/KappelSJ21}, and the output 
e.g., the reliability of the predictions \cite{DBLP:conf/icsoc/ComuzziCR18}. This will be reflected in an additional CMF on data properties and quality. 

\mypar{Conclusion} An additional requirement reflecting the control flow perspective of compliance constraints will be added to the CMF framework. The impact of concept drift on compliance predictions will be covered by the requirement to update the set of possible and future violations. Moreover, the CMF framework will be extended by a CMF on the ability to exploit external (process) context data and data properties and quality. These additional CMFs can be added to the new group \textsl{Data requirements}. Further on, a refinement of CMF8 and CMF9 will reflect the work on explaining and visualizing results of prediction. 

Note that the PPM approaches from the literature review will be analyzed with respect to how they meet PCM requirements in Section \ref{sec:prediction_requirements}. 

\subsection{PCM Literature Compilation and Findings}\label{subsec:slr_pcm}

For keywords and selection criteria used in the literature search for PCM see Fig. \ref{fig:lit_comp_wf}. 
The number of initial hits accounts to $510$ among which we selected papers based upon the outlined criteria additionally extended to include papers on SLA predictions in the context of service compositions. 
Papers from the medical domain predicting whether a medical treatment would result in the desired effects or whether patients are likely to follow the medical advice are considered out of scope due to a missing connection to business processes. Moreover, we excluded theses and finally came up with $7$ papers  \cite{leitner_data-driven_2013,DBLP:conf/sgai/KhanAAMCTN19,DBLP:journals/jowua/CicottiCDR15,DBLP:conf/icwe/RodriguezSDC10,DBLP:conf/sac/ComuzziCR19,DBLP:conf/icsoc/IvanovicCH11,DBLP:conf/icws/LeitnerMRD10}.
Out of these, \cite{leitner_data-driven_2013} was also found in the compilation of PPM literature and has therefore already been assessed.

PCM literature yields the following insights with respect to \textsl{i) existing CMFs from \cite{DBLP:journals/is/LyMMRA15}} and \textsl{ii) CMF extensions}. 
\cite{DBLP:conf/sgai/KhanAAMCTN19} can be classified as remaining time PPM approach
and hence addresses CMF1. \cite{DBLP:conf/sac/ComuzziCR19} mentions prediction across multiple process cases, i.e., instance spanning predictions, but no concrete solutions are provided. The approach is directed towards explainability by providing a measure for reliability of predictions, but just for individual cases. Those aspects are covered by CMF6 and CMF9.
\cite{DBLP:conf/icsoc/IvanovicCH11} uses an abstract notation for service orchestrations, i.e., \textsl{``compositions with a centralized control flow''} and \textsl{``predict possible situations of SLA conformance and violation, and to obtain information on the internal parameters of the orchestration (branch conditions, loop iterations) that may occur in these situation''}. These aspects are covered by CMF8 and CMF9 and touch distributed processes which was already identified as an additional CMF in the previous sections. \cite{DBLP:conf/icws/LeitnerMRD10} predicts SLAs and adapt service compositions in order to avoid a violation of SLAs. Mitigation actions and adaptations are part of CMF8. \cite{DBLP:journals/jowua/CicottiCDR15} targets the problem of state space explosion which addresses the newly added requirement on efficiency of compliance monitoring. An analysis of non-compliance to prevent compliance violations in the future with only limited prediction capabilities is presented in \cite{DBLP:conf/icwe/RodriguezSDC10} and addresses CMF8.

\mypar{Conclusion} PCM approaches confirm the findings of the CM and PPM literature reviews regarding CMF framework extensions.

\subsection{Comparison with Compliance Surveys}

The survey in \cite{DBLP:journals/kais/HashmiGLW18} outlines the following challenges and research gaps for compliance checking: \textsl{Formalisation of Norms, Norms Extraction and Elicitation, Multi-Jurisdictional Requirements, Control-Flow Structure, Integrating Rules with Processes, Handling Violations, Dealing with Model Evolution, Complexity and Performance} and \textsl{Usability and Generalisability} . From these challenges, \textsl{Control-Flow Structure} underpins our finding to add an explicit requirement for the control flow perspective. \textsl{Handling violations} as well as \textsl{Usability} correspond to user requirements CMF7--CMF10. \textsl{Model Evolution} is addressed by considering concept drift and in particular the effects of evolution and drift onto the compliance predictions. \textsl{Complexity and Performance} in \cite{DBLP:journals/kais/HashmiGLW18} mainly refers to the complexity of the compliance constraints which constitutes an important requirement for PCM, especially in calling for approaches where several process perspectives need to be predicted in combination. In this survey, complexity and performance additionally refers to the event streams gathered during runtime. The other challenges described in \cite{DBLP:journals/kais/HashmiGLW18} refer to the elicitation and formalization of compliance constraints which is outside the scope of this work. The survey presented in \cite{DBLP:journals/tsc/CabanillasRR22} analyzes compliance management frameworks along CMF1-CMF10 and adds three requirements, i.e., \textsl{R1:integration, R2:variability}, and \textsl{R3:reuse}. \textsl{R1:integration} corresponds to our finding to integrating event streams from multiple sources. \textsl{R2: variability} refers to the support of design and runtime compliance checks and \textsl{``provides explicit mechanisms to add an open-ended set
of checks''}. The latter underpins the importance of PCM as depicted in Fig. \ref{fig:overview_pcm_system}, i.e., supporting a number of (possibly combined) prediction goals, together with their impact on predicting compliance violations. \textsl{R3: reuse} is not covered by requirements in this work as we do not focus on maintaining the compliance constraint set. \cite{DBLP:journals/tsc/CabanillasRR22} mention compliance predictions as future work.

\subsection{Findings Based on Case Studies}\label{subsec:findings_casestudies}

Analogous to \cite{DBLP:journals/is/LyMMRA15}, we analyze case studies and real-world compliance constraint collections to identify CMFs with respect to
\textsl{i) existing CMFs \cite{DBLP:journals/is/LyMMRA15}} and \textsl{ii) possible CMF extensions}.  
Case studies can be found in various domains including data protection \cite{DBLP:journals/corr/abs-1811-03399}, finance \cite{DBLP:journals/bise/VoglhoferR20}, and manufacturing \cite{DBLP:conf/bpm/GallR21,DBLP:journals/is/WinterSR20}. As discussed in \cite{DBLP:journals/bise/VoglhoferR20}, real-world compliance constraints refer to the modeling requirements CMF1-3 plus control flow patterns such as existence, absence, and ordering. A collection of real-world constraints that span multiple process instances and processes can be found in \cite{DBLP:journals/corr/Rinderle-MaGFMI16}. Constraints spanning multiple instances are referred to by CMF6 in the original CMF framework \cite{DBLP:journals/is/LyMMRA15}. When looking into literature and the real-world constraints, CMF6 should address constraints that reflect i) the simultaneous execution of events, ii) constrained execution, iii) order of events, iv) non-concurrent execution of events, and v) constrained start of following instances \cite{DBLP:journals/is/WinterSR20,DBLP:conf/er/WinterR20}.

\mypar{Conclusion} Case studies and collections of real-world compliance constraints confirm modeling requirements CMF1-3. Additional requirements can be specifically identified in real-world constraints that span multiple process instances or processes and will be included in the assessment of CMF6. 

\subsection{Extended Compliance Monitoring Functionality Framework}\label{subsec:extended_cmf}

This section summarizes the extended CMF framework based on the findings from literature reviews and case studies in Sect. \ref{subsec:slr_cm} -- \ref{subsec:findings_casestudies}). We deliberately propose an extended CMF framework rather than, for example, a predictive CMF framework as the extensions refer to prediction and monitoring. Following \cite{DBLP:journals/is/LyMMRA15}, each of the CMFs is partitioned into sub CMFs reflecting specific requirements on the expressiveness of the CMF. 
Each of the CMFs 
is illustrated by an example in Sect. \ref{sec:prediction_requirements}.

The first extension refers to the modeling requirements by explication of CMF0 on control flow. Following control flow patterns for compliance constraints \cite{DBLP:journals/is/LyMMRA15}, we opt for the basic building blocks \textbf{existence (CMF0.1)}, \textbf{absence (CMF0.2)}, and \textbf{ordering (CMF0.3)}.
Note that CMF 1.1 time qualitative becomes obsolete due to adding CMF0.3 ordering. 

The user requirements are extended by refinement of CMF8 and CMF9. For CMF8, the \textbf{update of the set of possible and future violations (CMF8.3)} of compliance is added. CMF9 is renamed to \textbf{CMF9': Explainability}
and refined by \textbf{explain and visualize the prediction results (CMF9.3)}, \textbf{explain and visualize the set of possible and future violations (CMF9.4)}, and \textbf{explain and visualize the effects of mitigation actions (CMF9.5)}. Moreover, \textbf{CMF11 on efficiency/performance of CM} is added as user requirement.

Finally, the group of \textbf{Data requirements} on process event data such as logs and streams as input for solving the PCM problem is added. The consistency of the constraint base as also mentioned in the literature is considered beyond the scope of this work. In detail, the extensions comprise 
\textbf{integration of data from multiple sources (CMF12)}, \textbf{distributed processes (CMF13)}, \textbf{context data (CMF14)}, and \textbf{data properties and quality (CMF15)}.

\section{Extended CMF Framework, PCM System Requirements and Assessment}\label{sec:prediction_requirements}

In this section, we present the extended CMF framework in detail, identify PCM system requirements for each extended CMF, and analyze the $122$ papers from the PPM literature compilation plus the $24$ papers added by snowballing for their coverage of the extended CMF framework presented in Sect. \ref{subsec:extended_cmf}. 
 We structure the section along the extended CMFs beginning with the presentation of the CMFs and followed by the assessment for Sect. \ref{sub:modeling} to Sect. \ref{sub:data}. In the assessment, we discuss the relation between supporting CMF groups and their PCM system requirements and the prediction goal of the PPM approaches. The goal of the assessment is to understand the relation between CM, PPM, and PCM as well as to identify open research challenges (cf. Sect. \ref{sec:research_agenda}). 

The extended and updated CMFs when compared to the original framework presented in \cite{DBLP:journals/is/LyMMRA15} are described by means of the template also provided in \cite{DBLP:journals/is/LyMMRA15} in order to equip them with a more precise meaning.
The template contains for each CMF, its \textsl{name}, a brief \textsl{overview}, a \textsl{description}, \textsl{evaluation criteria} with particular focus on how the CMF could be verified through PPM in terms of PCM system requirements, \textsl{examples}, and clues on the \textsl{implementation} \cite{DBLP:journals/is/LyMMRA15}. In case of refined CMFs, the templates are centering around the refinements. Moreover, the original CMFs are also illustrated based on Ex. \ref{ex:eu} and the following constraints which apply to Ex. \ref{ex:eu} according to  \cite{european_parliament_regulation_2015,thomas_wenzel_transaction_2020,european_central_bank_reporting_2021,european_central_bank_mmsr_2017,european_banking_authority_guidelines_2018,bundesanstalt_fur_finanzdienstleistungsaufsicht_bankaufsichtliche_2021}.

Evaluation criteria are listed with the focus on PCM system requirements.
This illustration is understood as an addition to
the original CMFs presented in \cite{DBLP:journals/is/LyMMRA15}. The list of evaluation criteria results in a list of requirements for a holistic view that serves as an input for the assessment of existing PPM and CM approaches in meeting the criteria for a PCM system.

For the illustration and description of the extended CMF framework and elicitation of PCM system requirements, the distinction between predicate prediction and PCM (cf. Fig. \ref{fig:overview_pcm_system}) is important.
For predicate prediction, the expressiveness of the chosen logic for compliance constraints determines the assessment of an approach for modeling and execution requirement CMFs. For PCM, the assessment of modeling and execution requirement CMFs is also dependent on the prediction method. Therefore, the evaluation criteria for modeling and execution requirement CMFs in Sect. \ref{sub:modeling} and \ref{sub:uexec} focus on PCM system requirements for PCM. With the exception of CMF9.3 and CMF9.4 that coincide for predicate prediction, the user and data requirements are both similarly relevant to predicate prediction and PCM.

Furthermore, it is important to note that the prediction goal of an approach is independent of the ability to \textsl{match} reality observed through recorded/received events/activities with the respective events/activities stated in the compliance constraints. In predicate prediction, the approach must match before training the prediction method in order to compute the target for prediction, whereas in PCM the approach must match after the training, even after the prediction. Consequently, PCM separates the task of predicting from the task of matching and checking compliance. Furthermore, the ability of matching is required by any CM approach, i.e. also by approaches that do not predict compliance violations. Additionally note that matching requires an equivalence notion that formalizes the case when we consider two events/activities as equal. Since the often assumed label equivalence requires the usage of a controlled vocabulary throughout the design, execution and change of monitored processes and respective information systems, this equivalence notion is a non-valid simplification \cite{DBLP:journals/jwsr/Rinderle-MaRJ11}. Hence, further equivalence notions such as attribute equivalence are necessary for proper matching.  

Regarding the assessment of the extended CMFs and their PCM system requirements, we apply the following scale: $+$ indicates that the PCM system requirements for the CMF are fully met by existing PPM approaches, $\sim$ means the PCM system requirements are partly met and $-$ that the PCM system requirements are not met. We use $c$ to express that the requirements can be met by combining different existing PPM approaches, e.g., predicting resources is connected with predicting next activities. 

Note that we present an assessment that considers seen behavior, i.e., we assume that we have observed all behavior already in a historic log/stream. For the case of unseen behavior the assessment would for almost all CMFs evaluate to $-$ as only few approaches such as \cite{DBLP:conf/bpm/PauwelsC21} can deal with unseen next activity prediction in the form of updating the prediction model. Unseen behavior can be induced by changes and should not only be considered in the form of updates, but we could also aim at predicting unseen events based on, e.g., context data.

\subsection{Modeling Requirements}\label{sub:modeling}
CMF0 to CMF3 pertain to the ability of approaches to deal with control-flow, time, data and resource constraints.\\

\noindent\textbf{CMF0.x: existence, absence, ordering}

\textsl{Overview:} As basic building blocks of control-flow patterns, compliance rules need the ability to express that an activity must occur or exists in a process instance, that it is absent from a process instance and in what order two or more activities occur in the process instance.

\textsl{Description:} The existence or occurrence of an activity in a process instance can either be a primitive condition that a certain activity exists in the instance or be more advanced in the sense that the condition additionally carries information on how often the activity occurs in the instance. The former is stated either explicitly as a must occur or implicitly as part of another condition, for example, an activity data condition that implies the activity to occur. The latter typically adds some pattern for quantities such as "at least twice". To check conditions on absence of an activity in a process instance in particular requires the ability to decide whether a process instance is completed. The ordering conditions surpass the original time qualitative conditions in \cite{DBLP:journals/is/LyMMRA15} in expressiveness. Whereas time qualitative conditions can only express eventually follows, ordering conditions also capture conditions on the directly follows of events/activities or statements such as ``follows after three intermediate activities''.

\textsl{Evaluation criteria: } To fully support CMF0.x, an approach must have predicted the set of next activities/events ranked by probability of occurrence and with a distinction of directly/eventually follows or the complete order of activities/events per current process instance, the approach must have searched the complete process instance for activities/events and match recorded activities/events with the activity/event specified in the condition.

\textsl{Examples (cf. Ex. \ref{ex:eu})}: 
\begin{enumerate}
    \item[\namedlabel{ex_c1}{\textsl{C1}}] Activity ``Send secured MMSR report'' needs to be executed once per day (existence). 
    \item[\namedlabel{ex_c2}{\textsl{C2}}] Event ``Sending MMSR report failed'' by the authority must not occur (absence).
    \item[\namedlabel{ex_c3}{\textsl{C3}}] Activity ``Collect secured MMSR orders'' must be directly followed by ``Consolidate secured MMSR orders'' (ordering) .
\end{enumerate}

\textsl{Implementation:} For training the prediction method for predicate prediction or after prediction of the next activities/events for next activity prediction the approach must be able to check the truth of the existence and order conditions, i.e., it requires some equivalence notion for matching activities and events such as label equivalence or attribute equivalence \cite{koenig_compliance_2019} (e.g., in case events of conditions have to be matched with events contained in console logs that do not come with easily separable labels), a threshold for the probability of events/activities to decide whether they will actually occur, a mechanism to correlate events to process instances and a means to infer the order of the predicted events per process instance. For next activity prediction, the implementation needs to decide how to infer the absence of activities: Not predicting the occurrence of a certain activity is interpreted as being absent (only valid for approaches able to predict unseen behavior) or predicting it with a below threshold probability can be interpreted as being absent (valid interpretation for approaches able to predict unseen behavior and for approaches without that ability).

\noindent\textbf{CMF1: time quantitative}

\noindent For \textsl{Overview} and \textsl{Description}, and \textsl{Implementation} see \cite{DBLP:journals/is/LyMMRA15}. 

\textsl{Evaluation criteria:} To fully support the quantitative time functionality, the approach must have additionally predicted the remaining time to either complete the process, to complete the activity, the remaining time until the next event happens per process instance or timestamps for all predicted events/activities.

\textsl{Examples (cf. Ex. \ref{ex:eu}):} 
\begin{enumerate}
    \item[\namedlabel{ex_c4}{\textsl{C4}}] Activity ``Sending MMSR report succeeded'' needs to occur between $6$pm of the same day and before $7$am of the next trading day.
\end{enumerate}

\noindent\textbf{CMF2.x: activity data, case data}

\noindent For \textsl{Overview} and \textsl{Description}, and \textsl{Implementation} see \cite{DBLP:journals/is/LyMMRA15}. 

\textsl{Evaluation criteria:} To fully support these data-related functionalities, the approach must have additionally predicted relevant event attributes potentially depending on the prediction of case data over time.

\textsl{Examples (cf. Ex. \ref{ex:eu})}: 
\begin{enumerate}
    \item[\namedlabel{ex_c5}{\textsl{C5}}] Activity ``Correct order field entry'' with attributes ``reason=erroneous entry'' is to be sent as an amendment revision in the following report, i.e. the next ``Send secured MMSR report'' needs to include an XML amendment message for the respective order (constraint extracted from Fig. \ref{fig:constraint}).
\end{enumerate}

\noindent\textbf{CMF3.x: unary resource condition, extended resource condition}

\noindent For \textsl{Overview} and \textsl{Description}, and \textsl{Implementation} see \cite{DBLP:journals/is/LyMMRA15}. 

\textsl{Evaluation criteria:} To fully support resource-related functionalities, the approach must have additionally predicted the associated resource of activities/events based on resource predictions and potentially further event attributes necessary to monitor and check the conditions.

\textsl{Examples (cf. Ex. \ref{ex:eu})}: 
\begin{enumerate}
    \item[\namedlabel{ex_c6}{\textsl{C6}}] Orders without collateral need to be registered by a senior trader (unary resource condition).
    \item[\namedlabel{ex_c7}{\textsl{C7}}] Activities ``Send secured MMSR report'' can only be done by a role from a separate department than that of the role that initially registered the order (extended resource condition).
\end{enumerate}

\begin{table}[h!]
\begin{scriptsize}
\scalebox{1}{
\begin{tabular}{llc}
\toprule
Class & CMF    &  Coverage  \\
\midrule
\multirow{8}{*}{\makecell[l]{Modeling req.}} & \textbf{CMF0.1 existence}     & $+$ \\
 & \textbf{CMF0.2 absence}  & $+$ \\
 & \textbf{CMF0.3 ordering}  & $+$ \\
 & CMF1 time quantitative    & $c /+$\\
 & CMF2.1 activity data  & $+$\\
& CMF2.2 case data    & $\sim$ \\
& CMF3.1 unary resource condition  & $c /+$\\
& CMF3.2 extended resource condition & $c /+$ \\
\bottomrule\vspace{0.05em}
\end{tabular}
}
\centering
\caption{Coverage assessment of extended CMFs 1--3 by existing literature; CMF extensions in bold; $+$: covered, $\sim$: partly covered, $-$: not covered, $c$: combination necessary}\label{tab:extended_cmfs1-3}
\end{scriptsize}
\end{table}

\noindent\textbf{Assessment of modelling requirements:}

The following assessment of modelling requirements is summarized in Table \ref{tab:extended_cmfs1-3}.
The assessment of predicate prediction approaches with respect to modeling requirements in general depends on the expressiveness of the predicate language, i.e., which process perspectives (control flow, time, data, and resources) can be expressed as predicates. 
In the literature compilation, we identified $3$ distinct approaches that deal with predicting possible violations of predicates that range from simple SLAs \cite{DBLP:journals/dpd/LeitnerFHD13} to 
LTL based formulae \cite{DBLP:conf/caise/MaggiFDG14}, and first-order event expression (FOE) based formulae \cite{santoso_specification-driven_2020}. The predicates refer to control flow, time and event attributes. Hence, CMF0.1--0.3 are covered ($+$) if the activities have been already observed. \cite{DBLP:conf/caise/MaggiFDG14} is also able to deal with quantitative time prediction (CMF1). \cite{santoso_specification-driven_2020} explicitly deals with event attributes, i.e., activity data (CMF2.1, $+$), and resources (CMF3.x, $c/+$). Additionally, predicate prediction covers outcome-oriented PPM approaches that use case data as input for the prediction of outcomes \cite{lakshmanan_markov_2015}, but do not predict case data (CMF2.2, $\sim$).

Next activity / event prediction means to make statements about upcoming activities / events that are referred to by one or several compliance constraints. Take, for example, constraint \textsl{``If Correct order field entry activity is executed for a an order of the previous or even older day, then it must never be reported as a new order, but as an amendment.''} (cf. Ex. \ref{ex:eu}), which refers to activities \textsl{Correct order field entry} and \textsl{Send secured MMSR report}. Existing approaches predict next activities if the activities in the compliance constraint have already been observed so far. Absence of an activity can then also be implicitly predicted, based on probabilities. Consider for example a compliance constraint stating ``\textsl{$b$ must not directly occur after $a$}''. In this case, we would expect that for a trace in which we have observed $a$ the probability of $b$ as next activity should be 0. If it is not, we could end up with a compliance violation. Hence, next activity prediction approaches support CMF0.1, CMF0.2, and CMF0.3 ($+$).

Next activity / event prediction could also serve as ``anchor'' for predicting the modeling requirements CMF1--3.x referring to time, data, and resources by combining next event/activity prediction with remaining time and resource prediction. Except for \cite{DBLP:journals/access/RicoCDAG21} that predicts the future path of a trace and then the delay to the next event starting from the current event, combined approaches are missing. Instead time, data, and resources are used by existing approaches as features to improve next activity prediction. 

Finally, there are dedicated approaches for predicting remaining time, numeric indicators, and resources. 
Existing approaches for remaining time/delay in combination with next activity / event prediction cover CMF1 on quantitative time ($c/+$). In \cite{DBLP:journals/dss/ConfortiLRAH15}, activity data (CMF2.1) is used for prediction, i.e., cost for executing tasks is seen as a risk parameter.
Resource prediction (CMF3.1 and CMF3.2) in connection with temporal prediction is mostly seen from a scheduling perspective, i.e., how to determine and avoid potential temporal problems such as bottlenecks by assigning resources \cite{DBLP:journals/is/Rogge-SoltiW15,DBLP:journals/is/SenderovichWGM15}. Other approaches utilize resources as features for temporal predictions \cite{DBLP:conf/edoc/FolinoGP15,DBLP:journals/dss/KimCDMT22}.  \cite{DBLP:conf/bpm/0001DR19} predicts the resource / resource pool an upcoming event will be assigned to. Hence, CMF3.1 and CMF3.2 can be assessed with $c/+$.  

\subsection{Execution Requirements}\label{sub:uexec}
CMF4 to CMF6 pertain to the ability of approaches to deal with execution-based constraints. Furthermore, these CMFs enable the assessment of approaches to deal with domain-dependent information that is only available during execution of process instances.

\noindent\textbf{CMF4: non-atomic activities}

\noindent For \textsl{Overview} and \textsl{Description}, and \textsl{Implementation} see \cite{DBLP:journals/is/LyMMRA15}. 

\textsl{Evaluation criteria:} To fully support non-atomic activities, the approach must have additionally predicted and distinguished different event types and their relation to the respective activity.

\textsl{Examples (cf. Ex. \ref{ex:eu}):} 
\begin{enumerate}
    \item[\namedlabel{ex_c8}{\textsl{C8}}] A secured MMSR report can only be sent, if its processing is completed.
\end{enumerate}

\noindent\textbf{CMF5: life cycle}

\noindent For \textsl{Overview} and \textsl{Description}, and \textsl{Implementation} see \cite{DBLP:journals/is/LyMMRA15}. 

\textsl{Evaluation criteria:} To fully support life cycles, the approach must have additionally predicted and distinguished life cycle states/transitions of next events/activities.

\textsl{Examples (cf. Ex. \ref{ex:eu}):} 
\begin{enumerate}
    \item[\namedlabel{ex_c9}{\textsl{C9}}] A MMSR report processing starts with checking the MMSR sending channel, i.e. it puts the ``Send secured MMSR report'' activity into a ``ready'' state (activation). A successfully sent report event completes the corresponding ``Send secured MMSR report'' activity (activation+completion).
    \item[\namedlabel{ex_c10}{\textsl{C10}}] The sum of successfully sent reports and failed report sent events needs to equal the number of send report activations (balance start/comp\-l\-ete events). 
\end{enumerate}

\noindent\textbf{CMF6: Multiple instances constraints}

\textsl{Overview:} Multiple instances of compliance constraints may not only be necessary for a single process instance due to multiple occurrences of related activities, but in the case of instance-spanning constraints also for multiple instances of one or multiple process types.

\textsl{Description:} As constraints may impose requirements over multiple process instances or even processes and the constraint's instantiation trigger may occur multiple times during execution, each instantiation needs to be monitored simultaneously. Consider, for example, the instance-spanning constraint ``The centrifugation may only be started when at least five samples have arrived'' \cite{DBLP:conf/edoc/WinterR17}. If there exist five centrifuges in the lab, the instance-spanning constraint may be instantiated five times in parallel to monitor the arriving of samples at each centrifuge individually before the centrifugation may be started. If for four of the five centrifuges more than five samples have arrived and for the fifth centrifuge only three, but all centrifugations have started, the monitoring framework should only identify the respective constraint for the fifth centrifuge as violated.

\textsl{Evaluation criteria:} To fully support multiple instances functionalities, the approach must have predicted various combinations of previous prediction goals such as events with attributes, life cycles and/or timestamps in conjunction with a notion of process instances and processes to predict events across multiple instances/processes. In addition, the approach must have shown flexibility in handling various data granularities in constraints with respect to event attributes in received events, as the aggregation of data values for specific events may be necessary.

\textsl{Examples (cf. Ex. \ref{ex:eu}):} 
\begin{enumerate}
    \item[\namedlabel{ex_c11}{\textsl{C11}}] For each registered order, the order needs to be processed to be included in the final MMSR report (multiple instantiation).
    \item[\namedlabel{ex_c12}{\textsl{C12}}] All registered and processed orders are simultaneously sent to the authority (simultaneous execution).
    \item[\namedlabel{ex_c13}{\textsl{C13}}] All ``Process order'' activities must be completed by $7$am of the following trade day (constrained execution).
\item[\namedlabel{ex_c14}{\textsl{C14}}] Each report needs to be sent without another send operation in concurrence (non-concurrent activities). 
\item[\namedlabel{ex_c15}{\textsl{C15}}] Only four MMSR reports can be sent to the ECB per day (constrained start).
\end{enumerate}

\textsl{Implementation:} Encoding inter-case features \cite{DBLP:journals/is/SenderovichFM19} can be a solution to predict, for example, the utilization of shared resources. Compliance prediction of constraints that span across multiple instances or processes pose requirements on the ability of a PCM solution to a) model \textsl{context} in the constraints, e.g., the set of instances the constraint refers to  \cite{DBLP:conf/bpm/FdhilaGRMI16}; b) differentiate process instances/processes; c) track the contexts of rule activations and their compliance status separately; d) transfer the separate handling of rule activations and their contexts to instance-spanning constraints.

\begin{table}[h!]
\begin{scriptsize}
\scalebox{1}{
\begin{tabular}{llc}
\toprule
Class & CMF    &  Coverage  \\
\midrule
\multirow{3}{*}{\makecell[l]{Execution req.}} &  CMF4 non-atomic activities  & $-$ \\
& CMF5 life cycle &  $-$\\
& CMF6 multiple instances constraints & $c /\sim$ \\
\bottomrule\vspace{0.05em}
\end{tabular}
}
\centering
\caption{Coverage assessment of extended CMFs 4-6 by existing literature; CMF extensions in bold; $+$: covered, $\sim$: partly covered, $-$: not covered, $c$: combination necessary}\label{tab:extended_cmfs4-6}
\end{scriptsize}
\end{table}

\noindent\textbf{Assessment of execution requirements CMF4--CMF6:}
The following assessment of execution requirements is summarized in Table \ref{tab:extended_cmfs4-6}.
If the input event stream contains different event types (CMF4, CMF5) such as \texttt{start}, \texttt{complete}, or \texttt{running} (cf. life cycle model for XES \cite{7740858}), the corresponding event labels are conceptually encoded as features, but a demonstration of correctly handling lifecycle event labels as features in case of predicate prediction and, additionally, predicting lifecycle event labels for future events in case of PCM is missing. 

For CMF6, based on the case studies outlined in Sect. \ref{sec:literature_compilation} we can provide a very detailed level of assessment which is not reflected in the CMF due to the sake of granularity levels among the CMFs. In particular, the assessment of CMF6 on multiple instance constraints, will follow the categorization for instance-spanning constraints proposed in \cite{DBLP:journals/is/WinterSR20}. Existing PPM approaches to address the multiple instantiation of compliance constraints are missing. For the simultaneous execution of process instances, inter-case features for batching (i.e., executing process instances in one batch) are used in order to improve remaining time predictions \cite{DBLP:journals/is/SenderovichFM19,DBLP:conf/icpm/KlijnF20,DBLP:conf/icpm/PourbafraniKKA21}. Aggregated PPIs that might include data constraints can be predicted based on \cite{DBLP:conf/ideas/CuzzocreaFGP18} as well as aggregated risks over multiple instances by \cite{DBLP:journals/dss/ConfortiLRAH15}. 
\cite{firouzian_real-time_2019} propose a probabilistic approach for remaining time prediction taking into account hidden dependencies between process instances. Constrained execution of instances as well as non-concurrent execution of instances are not covered by temporal prediction approaches. Constrained start of instances is only touched upon, i.e., \cite{DBLP:conf/edoc/FolinoGP15} predict how many instances will start in a particular time window.

\subsection{User Requirements}\label{sub:user}

CMF7 to CMF11 pertain to the ability of approaches to support the user in understanding and managing compliance violations in time. 

\noindent\textbf{CMF7: reactive management}

\noindent For \textsl{Overview} and \textsl{Description}, and \textsl{Implementation} see \cite{DBLP:journals/is/LyMMRA15}.

\textsl{Evaluation criteria:} To fully support the reactive management of compliance violations, the approach must have continued predicting events for process instances that have already violated constraints and to have continued the monitoring of further constraints for these instances. On the one hand, continuous monitoring of violations gives a more complete and fine-grained view on the compliance status of a process/system. On the other hand, it supports and enriches further functionalities of reactive management such as recovery, compensation mechanisms and reporting/documentation features for auditing.

\textsl{Examples (cf. Ex. \ref{ex:eu}):}  
\begin{enumerate}
    
\item[\namedlabel{ex_c16}{\textsl{C16}}] Unplanned deviations from normal operations (disruptions) and their root causes have to be registered, evaluated and prioritized with respect to their resulting risks and escalated according to predetermined criteria (continuous monitoring).
\end{enumerate}

\noindent\textbf{CMF8.x: early detection of conflicting rules, possible/future violation, update set of possible and future violations, recommendations for users to avoid violations}

\textsl{Overview:} The ability for pro-active detection and management of compliance violations does not only include the detection of conflicting rules that lead to implicit violations, the detection of possible and inevitable future violations and subsequent recommendation of mitigation actions, but also includes the capacity of an approach to react to evolving or high velocity processes in real-world environments by updating the set of possible and future violations.

\textsl{Description:} The typical training of a prediction model on ex-post process instances may be too inflexible to capture the dynamics and variability of evolving processes in the real-world \cite{DBLP:journals/corr/abs-2109-03501}. Hence, approaches should be able to flexibly deal with processes exhibiting concept drift or a multitude of variants over time through mechanisms to update the prediction model and/or the set of possible and future violations as the process evolves.

\textsl{Evaluation criteria:} To fully support the pro-active detection and management of compliance violations, the approach must have detected conflicting rules as soon as possible with precise probability/likelihood, to continuously update the set of conflicting rules as the event stream evolves, to detect and predict the set of compliance violations as soon as possible and as complete as possible with precise probability/likelihood and to continuously update compliance prediction for all compliance constraints and events as the event stream evolves. Furthermore, the approach has to determine and provide mitigation actions based on compliance predictions as soon as possible, with precise assessment of risk and impact of the mitigation actions. Additionally, it has to continuously update the mitigation actions based on updates of compliance predictions.

\textsl{Examples (cf. Ex. \ref{ex:eu}):} \\
$\bullet$ Consider \textsl{Scenario \ref{sce:rules}}. The accuracy constraint of the MMRS regulation is in conflict with a constraint set by the German ``Arbeitszeitgesetz'' (early detection of conflicting rules); \\ $\bullet$ Consider \textsl{Scenario \ref{sce:orders}}. Missing the deadline for sending the report is a possible violation to be predicted (possible/future violation); \\ $\bullet$ Consider \textsl{Scenario \ref{sce:orders}}. Each time an activity of the data collection and consolidation in the staging area is completed, the likelihood of possibly violating the MMSR timeliness constraint may change (update set of possible and future violations); \\ $\bullet$ Consider \textsl{Scenario \ref{sce:orders}}. Recommendations to avoid the possible violation may be to provision additional server resources and register them with the execution engine in the staging area; or to avoid all future possible violations may be to reduce the number of correcting order field entries by putting attention to the topic in the front-office or by aligning their incentives with a correcting order field activity metric (recommendations for users);

\textsl{Implementation:} For updating the set of possible and future violations as the event stream evolves, the CM system can either take a brute-force approach by always updating the predictions as new events arrive, which may be computationally intensive or even infeasible for high velocity streams, or optimize the points in time to update the predictions as well as the prediction model according to new events. Consequently, the approach needs both a strategy for updating the prediction and for updating the prediction model in light of concept drift. These strategies can either be optimized for the particularities of the overall PCM system or generic and configurable for various settings.

\noindent\textbf{CMF9.x: root cause analysis, effective communication of root cause, explain and visualize prediction results, explain and visualize set of possible and future violations, explain and visualize effects of mitigation actions}

\textsl{Overview:} Next to root cause analysis and the effective communication of identified root causes, the prediction results, set of possible and future violations and the mitigation effects need to be communicated to the end user in a meaningful way by appropriate visualizations and explanatory presentations. These additional abilities increase the usability of the approach and facilitate the efficient use of the same in practice.

\textsl{Description:} Depending on the prediction goal(s) of the approach, the prediction results and the set of possible and future violations may coincide (in case of predicting truth values of constraints) or be separate from another (in case of predicting the sequence of events before evaluating the constraints on the sequence of events). Nevertheless, for both and the effects of mitigation actions, the approach should provide the user with explanations and visualizations in support of a quick assimilation of the key factors determining the results, what they mean and - for mitigation actions - what their execution will lead to.

\textsl{Evaluation criteria:} To fully support root cause analysis, explainability and visualization functionalities, the approach must have precisely determined root causes for predicted compliance violations as soon as possible, provided root cause analysis and continuously visualized root causes for predicted compliance violations to users. Furthermore, the approach has to have continuously provided explanations for compliance predictions at algorithmic level (i.e., which input leads to which output) and continuously visualized the prediction results in their context, possibly together with providing post hoc explanations (together with CMF9.1). The predicted compliance violations have to have be continuously visualized together with their root causes and effects (cf. CMF9.1/CMF9.3) and with their mitigation actions and the effects of applying the mitigation actions. Each of these functionalities requires the demonstrated ability to handle single and multiple instances (the latter also in an aggregated manner) and multiple process perspectives and views.

\textsl{Examples (cf. Ex. \ref{ex:eu}):} \\
\noindent$\bullet$ Consider \textsl{Scenario \ref{sce:orders}}. The root cause analysis for the longer Middle- and Back-office activities should reveal the exceptionally high number of correct order activities as the root cause (root cause analysis);\\
$\bullet$ Consider \textsl{Scenario \ref{sce:orders}}. Effective communication of the exceptionally high number of correct order activities both entails a visualization for each MMSR reporting process showing the relation of the correct order activities to the respective, longer running activities in the Middle- and Back-Office with the respective correlations and statistical tests (effective communication);\\
$\bullet$ Consider \textsl{Scenario \ref{sce:orders}}. The prediction of the later completion times for the Staging area should explain that this is due to the longer running Middle- and Back-Office activities and to what extent this is the case. The quantifying metrics for the extent are visualized in a graph showing all the features and to what extent they determine the current prediction and their likelihood (prediction results);\\
 $\bullet$ Consider \textsl{Scenario \ref{sce:orders}}. The content of CMF9.2 in conjunction with a visualization of the predicted overtime by which the staging area will violate the timeliness requirement assembles all required information for the MMSR reporter to understand the possible violation and decide on how to proceed (set of possible/future violations);\\
 $\bullet$ Consider \textsl{Scenario \ref{sce:orders}}. The type of provisioned servers and their effect is visualized to show the to what extent they affect the performance of the staging area. The properties of the execution engine that determine the effect and its limits is explained on top of the visualization (mitigation actions);

\textsl{Implementation:} To streamline the result presentation of the root cause analysis, prediction (and compliance checking) and mitigation actions with their respective effect to the end user, the implementation can choose to support tools for \textsl{Business Intelligence} \cite{grossmann_fundamentals_2015} or implement a standalone graphical user interface. The former option transfers the functionalities with respect to communicating, explaining and visualizing (CMF9.2--CMF9.5) to an external tool and the end user. Since the end user has to know the technical details of acquiring relevant information from the approach, this option may entail a considerable barrier for using the system. The latter option does not come with these limitations. Here, the approach can leverage \textsl{eXplainable Artificial Intelligence} \cite{burkart_survey_2021} to support explainability and communication of results and \textsl{Visual Analytics} \cite{thomas_visual_2006} to enable the interaction of the end user with the results (CMF9.2--CMF9.5). By summarizing the key results and their implications in a dashboard (as proposed, for example, in \cite{hutchison_design_2010}), the approach can support the user in understanding the various outputs of the system, as presenting information through dashboards has proven beneficial in practice \cite{eckerson2010performance,khatuwal_business_2022,kumar_performance_2017}.\\

\noindent\textbf{CMF10.x: compliance degree of single traces, compliance degree of an entire process/system}

\noindent For \textsl{Overview} and \textsl{Description}, and \textsl{Implementation} see \cite{DBLP:journals/is/LyMMRA15}. 

\textsl{Evaluation criteria:} To fully support compliance degree functionalities, the approach must have continuously exploited the predicted probabilities/likeli\-hoods of compliance violations for continuously determining and updating the compliance degree of single process instances and across all process instances and processes. Depending on the complexity of the compliance constraint base, the prediction goal in the case of predicate prediction can be tweaked to predict compliance degree as the probability of non-violation instead of truth values for compliance constraints. Note that several compliance constraints can apply to the same case which can result in one compliance constraint being predicted to be violated but the others being predicted to be fulfilled. A compliance degree needs to take this into account and eventually also consider the severeness of a potential violation in particular when considering CMF8.x and CMF9.x. Maybe violating one compliance constraint that has not such severe consequences can be taken into account in order to improve the overall compliance degree either within one trace or even for the entire system.

\textsl{Examples (cf. Ex. \ref{ex:eu}):} \\$\bullet$ The respective regulations are different for the four reporting processes and the institution is required to monitor and explain deviations in the reporting processes leading to the need for individual compliance degrees of each reporting process instance (single traces); \\ $\bullet$ To assess the overall risk that is imposed through the transaction-based reporting system, the bank must measure and monitor the compliance degree of the entire process/system \cite{bundesanstalt_fur_finanzdienstleistungsaufsicht_bankaufsichtliche_2021} (entire process/system);

\noindent\textbf{CMF11: efficiency/performance of CM}

\textsl{Overview:} To realize pro-active detection and management of compliance violations, an appropriate performance of the approach is a prerequisite.

\textsl{Description:} As the efficiency/performance of the approach determines the point in time the end user can react to compliance violations by executing proposed compensation/recovery mechanisms, reporting the violation and documenting it; or proactively act to mitigate the possible violations by executing proposed mitigation actions, it is crucial that this point in time adheres to the performance requirements set for the monitored system and it comes before the possible violations occurs. Depending on how challenging the performance requirements are, a particular approach may be unsuitable.

\textsl{Evaluation criteria:} To fully support the performance functionality, the approach must have provided performance optimization strategies for compliance prediction and its continuous update based on, e.g., delta approaches. Furthermore, it needs to have provided benchmarks with respect to compliance prediction performance in offline and online settings.

\textsl{Examples (cf. Ex. \ref{ex:eu}):} If the approach takes more than 11 hours to update its predictions, it can never capture the dynamics inherent to the reporting process at runtime, rendering its capabilities as only reactive or for auditing purposes; 

\textsl{Implementation:} Due to the trade-off between performance and accuracy \cite{sidiroglou-douskos_managing_2011}, approaches with the highest accuracy values in offline settings may render themselves unsuitable in online settings. As performance requirements on the approach may greatly vary, the implementation can either specialize on a certain performance threshold in online settings and optimize its results accordingly or introduce configuration parameters/options that determine the effective performance/efficiency of the approach in practice.

\begin{table}[h!]
\begin{scriptsize}
\scalebox{0.94}{
\begin{tabular}{llc}
\toprule
Class & CMF    &  Coverage  \\
\midrule
\multirow{13}{*}{\makecell[l]{User req.}} & CMF7 reactive management &  + \\
& \makecell[l]{CMF8.1 early detection of conflicting rules} &  $+$\\
& CMF8.2 possible/future violation &  $+$\\
& \makecell[l]{\textbf{CMF8.3 update set of possible and future violations}} &  $c /\sim$ \\
& \makecell[l]{CMF8.4 recommendations for users to avoid violations} &  $\sim$ \\
& CMF9.1 root cause analysis &  $c /\sim$ \\
& \makecell[l]{CMF9.2 effective communication of root cause} &  $-$ \\
& \makecell[l]{\textbf{CMF9.3 explain and visualize prediction results}} &  $c /\sim$ \\
& \makecell[l]{\textbf{CMF9.4 explain and visualize set of possible and future violations}} &  $\sim$ \\
& \makecell[l]{\textbf{CMF9.5 explain and visualize effects of mitigation actions}} &  $-$ \\
& \makecell[l]{CMF10.1 compliance degree of single traces} &  $-$\\
& \makecell[l]{CMF10.2 compliance degree of an entire process/system} &  $-$\\
& \makecell[l]{CMF11 efficiency/performance of CM} &  $\sim$ \\
\bottomrule\vspace{0.05em}
\end{tabular}
}
\centering
\caption{Coverage assessment of extended CMFs 7--11 by existing literature; CMF extensions in bold; $+$: covered, $\sim$: partly covered, $-$: not covered, $c$: combination necessary}\label{tab:extended_cmfs7-11}
\end{scriptsize}
\end{table}

\noindent\textbf{Assessment of user requirements CMF7--CMF11:}

The following assessment of user requirements is summarized in Table \ref{tab:extended_cmfs7-11}.
In general, it is often difficult or impossible to identify the output of existing prediction approaches. However, the PCM system requirements state to at least predict the set of next activities / events, ranked by probability of occurrence and a distinction between immediately/eventually occurs for PCM. If we look at more complex compliance constraints referring to several activities and their occurrence/absence and order, possibly in combination with time, data, and resources, a fine-granular prediction feedback with probabilities would be desired both for predicate prediction and PCM, which is basically possible, but not explicitly provided by any of the approaches. 

Regarding CMF7, prescriptive monitoring approaches \cite{DBLP:conf/icpm/BozorgiTDRP21,DBLP:journals/kais/Fahrenkrog-Petersen22} constitute a means to propose demonstrated recovery or compensation in the form of interventions to the user and provide feedback ($+$).

Regarding CMF8.1 and CMF8.2, predicate prediction approaches  \cite{DBLP:conf/caise/MaggiFDG14,DBLP:journals/dpd/LeitnerFHD13,santoso_specification-driven_2020} can basically provide early detection of (future) compliance violations ($+$). Updates of prediction results, especially compliance violations (CMF8.3), and prediction models is addressed in a preliminary way by incremental learning approaches that focus on updating the prediction model when drifts occur, e.g., \cite{DBLP:conf/bpm/PauwelsC21,DBLP:conf/IEEEscc/MaisenbacherW17} as well as based on continuous task monitoring through sliding windows as proposed in, e.g., \cite{DBLP:conf/ideas/CuzzocreaFGP18} ($c/\sim$). 
For CMF8.4, approaches mention that predictions can provide recommendations for users \cite{DBLP:conf/caise/MaggiFDG14}. Alerts \cite{DBLP:conf/bpm/CabanillasCMB14} and dashboards \cite{DBLP:journals/computing/PolatoSBL18,DBLP:conf/icseng/FerreiraSAH14,hutchison_design_2010} provide information on predictions and compliance to users and can help to avoid violations. Risk predictions \cite{DBLP:journals/dss/ConfortiLRAH15} are provided to users as recommendations (CMF8.4) which can also partly serve as mitigation actions for lowering risk for specific risk types. However, all of these recommendations do not target compliance violations. Prescriptive monitoring approaches can foster the early detection of compliance violations (CMF8.1) and the preparation of mitigation actions (CMF8.4). \cite{DBLP:conf/icpm/BozorgiTDRP21,DBLP:conf/icpm/ShoushD21,DBLP:journals/kais/Fahrenkrog-Petersen22}, for example, enable the generation of alarms that trigger interventions to prevent an undesired outcome or mitigate its effect. Similarly, \cite{DBLP:conf/ecis/WeinzierlDTZM21} \textsl{``supports the proactive handling of deviations, i.e. inserted and missing events in process instances, to reduce their potential harm''}. However, those approaches do only provide a limited set of recommendations or mitigation action, in particular, with respect to a set of constraints and not KPIs or SLAs. Yet, for a PCM system, approaches need to provide an extensive set of interventions together with an evaluation of their effects, i.e., since multiple constraints can be in effect at the same time, mitigation actions need to consider the interplay of those constraints as well. ($\sim$ for CMF8.4). 

Root cause analysis (CMF9.1) can be implicitly based on probabilities and feature vectors, i.e., by answering the question whether certain data elements influence the prediction of the next activity/event \cite{DBLP:journals/kbs/JalayerKPB22,galanti_explainable_2020,doi:10.1080/12460125.2020.1780780} ($c/\sim$). However, the effective communication of root causes to users (CMF9.2) is missing ($-$), although explainability (CMF9.3), e.g., based on features, is targeted by several approaches recently \cite{DBLP:conf/ecis/StierleBWZM021,galanti_explainable_2020,DBLP:conf/ideas/CuzzocreaFGP18,Mehdiyev2021,doi:10.1080/12460125.2020.1780780} ($c/\sim$). CMF9.3 on explaining and visualizing results is implicitly supported via helping to choose parameters by \cite{DBLP:conf/caise/BartmannHCDD21} and by \cite{DBLP:conf/icpm/KlijnF20} in the context of inter-case features for batching. Quality metrics for the prediction results are provided, including stability \cite{DBLP:journals/access/KimC21e} and reliability \cite{DBLP:conf/icsoc/ComuzziCR18}. However, existing approaches lack the ability to explain the prediction results in natural language and by means of visualizations that both are comprehensible by domain experts without knowledge of the prediction method or without a mathematical background ($c/\sim$). CMF9.4 on explaining and visualizing predicted violations is supported partly by \cite{galanti_explainable_2020} that visualize and explain predicted violations technically ($\sim$), i.e. not in natural language similar to the previous explanation and visualization of prediction results. In particular, visualization approaches for explaining prediction results and the effects of mitigation actions (CMF9.5) are missing ($-$). 

For the assessment of compliance degrees, there is no approach for single instances (CMF10.1, $-$) and approaches like \cite{DBLP:journals/is/SenderovichFM19} encode inter-case features but do not provide means for predicting them, i.e., CMF10.2 is assessed as $-$. 

Regarding efficiency and performance (CMF11), first approaches contribute by applying, for example, scalable online learning algorithms, \cite{DBLP:journals/access/RicoCDAG21}, hyperparameter optimization \cite{DBLP:journals/is/Francescomarino18}, and temporal predictions \textsl{``in a parallel and distributed manner, on top of a cloud-based service-oriented infrastructure''} \cite{DBLP:conf/IEEEares/CesarioFGP16}, yet fail to develop a case study that elicits user requirements on the performance/efficiency of the system and comprehensively benchmarks existing approaches based on the case study; only for outcome-oriented approaches a benchmark exists for execution times \cite{teinemaa_outcome-oriented_2019} ($\sim$).

\subsection{Data Requirements}\label{sub:data}
CMF12 to CMF15 pertain to the ability of approaches to deal with properties of data that are determined by the means of recording and storing the data and the properties of data sources.

\noindent\textbf{CMF12: integration of data from multiple sources} 

\textsl{Overview:} Real-world processes can be supported by multiple information systems at once thereby requiring the approach to integrate the data from each of the information systems to construct the complete processes. If context data is considered, it needs to also be integrated with the event data.

\textsl{Description:} When business processes are supported by multiple, distributed information systems, observing events by extracting them from the various information systems as a starting point for approaches can become challenging, since there exists a high degree of heterogeneity among the systems \cite{niedermaier_observability_2019} leading to a need for subsequent integration. Reasons for companies to support their processes with multiple information systems can be specialized software or external systems that support the process, e.g. after outsourcing \cite{kakabadse2005outsourcing}. If these information systems do not follow the same data governance, integration becomes inevitable \cite{castro_ontological-based_2021}.

\textsl{Evaluation criteria:} To fully support the data integration functionality, the approach must have supported various event extraction techniques and have transitioned from basing predictions on label equivalence to equivalence notions based on activity semantics, e.g., attribute equivalence \cite{DBLP:journals/jwsr/Rinderle-MaRJ11} and integration of other process perspectives and case ids.

\textsl{Examples (cf. Ex. \ref{ex:eu}):}\\
$\bullet$ The activities of the transaction reporting process can either be explicitly recorded through the underlying events in information systems such as Murex\footnote{https://www.murex.com/} or Bloomberg\footnote{https://www.bloomberg.com/} or be implicitly recorded as database transactions in ERP systems such as SAP\footnote{https://www.sap.com/}. The various data sources together with additional data sources for context data have to be integrated.\\
$\bullet$ ``If the loan request is greater or equal to one million, the solvency level of the customer needs to be at least A, a manager needs to process the request, and the solvency information must not be older than two days. [...] the information necessary to check this rule is distributed across multiple systems \cite{koenig_compliance_2019}.

\textsl{Implementation:} The approach can address the challenge of multiple, potentially heterogeneous sources by supporting various event extraction techniques, equivalence notions and event abstraction mechanisms.\\

\noindent\textbf{CMF13: distributed processes}

\textsl{Overview:} Compliance in process choreographies refers to the fulfillment or violation of constraints at different levels, i.e., the global choreography level, the local private process level, a mixture of both, and assertions \cite{fdhila2022verifying}, i.e., constraints might span across multiple partners of the choreography. In addition, constraints might refer to one or several choreography instances (and span across multiple partners at the same time).

\textsl{Description:} 
Constraints in distributed settings span across several partners in a process choreography and additionally might span one or several choreography instances. Due to the distribution of the partners, the events of process choreographies are typically recorded in multiple information systems. Hence, CMF13 is related to CMF12 and predicting compliance in distributed settings is a task that typically cannot be performed at one partner's side, but across several partners in the choreography. Additionally, compliance prediction has to deal with confidentiality issues if private processes of partners are affected. Overall, dealing with process choreographies poses new challenges for the approach, e.g. the prediction of and checking of constraints under privacy and confidentiality issues \cite{fdhila2022verifying} and the realization of the PCM system with respect to how the prediction is actually performed (at one partner's side, in a distributed way?).  

\textsl{Evaluation criteria:} To fully support the distributed processes functionality, the approach must have transitioned from basing predictions on label equivalence to equivalence notions based on activity semantics, e.g., attribute equivalence \cite{DBLP:journals/jwsr/Rinderle-MaRJ11} and integration of other process perspectives, case ids, and message ids. Additionally, the approach must have provided compliance predictions on the event streams/compliance constraints with confidentiality requirements constituted by, e.g., hidden private process information.

\textsl{Examples (cf. Ex. \ref{ex:eu}):}  All eight reporting processes run concurrently. To assess and predict the overall report data collection and consolidation performance and completeness for internal reporting purposes, equivalence notions based on activity semantics have to be employed.

\textsl{Implementation:} So far only few approaches have addressed the challenges that come with compliance in process choreographies. In \cite{fdhila2022verifying}, a decomposition algorithm is suggested to enable the distributed and confidential verification of compliance constraints that spread over several process participants as private information is common in competitive markets. For realizing and implementing distributed compliance verification mechanisms, technologies such as blockchains \cite{DBLP:journals/tmis/MendlingWABCDDC18} could be used. The other way round, process event logs/streams could be collected from the partners and merged where partners with private information provide abstracted log information \cite{DBLP:conf/smds/Aalst21}. In this case, compliance verification can be realized in a centralized way using existing frameworks and implementations, but it has to be investigated how abstraction influences compliance verification quality. Approaches for predicting compliance in process choreographies are missing. 

\noindent\textbf{CMF14: context data}

\textsl{Overview:} Context data, i.e., data that is internal or external to the process data, has proven useful in enabling or enriching CMFs such as root cause analysis and explainability (cf. Sect. \ref{sec:literature_compilation}), so approaches should come with the ability to deal with such data.

\textsl{Description:} Taking context data for pro-active detection and management of compliance violations into account includes the ability to identify relevant context data, relating the context data to the events/activities that are actually affected or in that context represented by the data, leveraging context data for improving predictions (e.g. potentially enabling predictions for processes whose events are private) and considering them as further evidence for root causes and for explaining what the compliance violation means. In the case of external context data, it is likely that this data is not recorded in the same information system as the process, which relates this functionality with CMF12.

\textsl{Evaluation criteria:} To fully support the data integration functionality, the approach must have continuously exploited context data for compliance predictions, particularly for predictions at the presence of unseen process and data behavior and for predicting unseen context data behavior. Furthermore, the approach must have continuously exploited context information for explaining prediction results.

\textsl{Examples (cf. Ex. \ref{ex:eu}):} In the case of major external events such as the beginning of the Ucraine war in 2022, trading activity in financial markets can be higher than usual leading to an increase number of orders that need to be reported, putting a significantly higher burden on the reporting process (cf. Ex. \ref{ex:eu}). In this case, the prediction of timestamps and durations is affected by the context of the Ucraine war that affects the external trading activity in financial markets measured as trading volume for all markets and therefore, the possible violation of the timeliness constraints for the internal processes.

\textsl{Implementation:} For each ability of dealing with context data, there exist proposed approaches, e.g. context ontologies for processes can guide the identification of relevant context \cite{vanderaalst2012context,costa2007contextrules,Song2019contextreview,Kronsbein2014contextinternal-org-resource-customer-external,park2022detecting}. Hence, the approach may either connect existing approaches in a meaningful way or opt to develop new methods for each required step.

\noindent\textbf{CMF15: data properties and quality}

\textsl{Overview:} Properties of the process and context data may require data transformations or suitable relation mechanisms (cf. CMF 12) as part of data preprocessing \cite{famili1997data}, whereas the quality of data can lead to the exclusion of data attributes during preprocessing. In the context of CM, excluding data for quality reasons may limit the monitoring to events of process instances that are more likely to be compliant.

\textsl{Description:} 
For CM, a crucial data property is transparency, i.e., to understand which properties of data lead to which effects during execution of the approach to facilitate the understanding of the results (cf. CMF9.x). As the goal of CM is to reactively and pro-actively detect and manage compliance violations, events with data quality issues can signify compliance violation. Consequently, approaches should avoid preprocessing to improve data quality, in particular during runtime, but rather have the ability to flexibly deal with the recorded data quality.

\textsl{Evaluation criteria:} To fully support the data properties/quality functionality, the approach must have considered and exploited properties and quality of the input event streams, interpreted data (quality) properties with respect to prediction results and elaborated strategies for dealing with data quality properties and problems under prediction result quality guarantees.

\textsl{Examples:}\\
$\bullet$ Since data quality issues such as erroneous values or wrong formats either lead to more "Correct order field entry" activities or to amendment/correction messages to be included in the next report \cite{european_central_bank_reporting_2021}, they affect both what activities in the reporting process occur and their performance.\\
$\bullet$ ``For example, although the average number of activities in the EnvLog dataset is only 44 (compared to 20 for BPI’12), the dataset only provides 787 instances for 331 possible activities resulting in a high sparsity of 0.42, whereas BPI’12 has a comparably low sparsity of 0.0028.'' \cite{DBLP:journals/dss/HeinrichZJB21}.

\textsl{Implementation:} So far research has yet to address how to exploit data properties and quality for CM.

\begin{table}[h!]
\begin{scriptsize}
\scalebox{1}{
\begin{tabular}{llc}
\toprule
Class & CMF    &  Coverage  \\
\midrule
\multirow{4}{*}{\makecell[l]{Data req.}} & \makecell[l]{\textbf{CMF12 integration of data from multiple sources}} &  $\sim$ \\
& \textbf{CMF13 distributed processes} &  $-$ \\
& \textbf{CMF14 context data (internal $|$ external)} &  $c /+$ $|$ $c/\sim$ \\
& \textbf{CMF15 data properties and quality} &  $\sim$ \\
\bottomrule\vspace{0.05em}
\end{tabular}
}
\centering
\caption{Coverage assessment of extended CMFs 12--15 by existing literature; CMF extensions in bold; $+$: covered, $\sim$: partly covered, $-$: not covered, $c$: combination necessary}\label{tab:extended_cmfs12-15}
\end{scriptsize}
\end{table}

\noindent\textbf{Assessment of data requirements CMF12--CMF15:}

The following assessment of data requirements is summarized in Table \ref{tab:extended_cmfs12-15}. It does not distinguish between predicate prediction and PCM, as the approach does not affect the assessment of PCM system data requirements.

For CMF12 on the integration of data from multiple sources, \cite{DBLP:conf/bis/PegoraroUGA21,DBLP:conf/bpm/TeinemaaDMF16} deal with structured and unstructured data, i.e., textual data, as input for PPM ($\sim$). There are no approaches for predictions in distributed processes (CMF13, $-$). 

The potential of context data (CMF14) is mentioned by several approaches. Internal context data is exploited by existing approaches such as \cite{DBLP:journals/is/BrunkSPRMB21,DBLP:journals/access/ChamorroRRdSR20,DBLP:journals/kbs/JalayerKPB22,DBLP:conf/otm/YeshchenkoDRMS18} by encoding them as features for next activity/event prediction. \cite{DBLP:journals/access/ChamorroRRdSR20} gather domain expert feedback on the selection of the context data. Internal context data is also utilized by \cite{DBLP:conf/otm/FolinoGP12} for temporal predictions ($c/+$). \cite{DBLP:conf/otm/YeshchenkoDRMS18} include information from news articles as context data into the predictive process monitoring based on sentiment analysis, which contributes to the exploitation of external context data as one possible sources ($c/\sim$). Further sources for external data comprise, e.g., sensor data streams \cite{DBLP:conf/bpm/StertzRM20}.

The influence of data quality and properties (CMF15) is considered by first approaches that consider data properties \cite{DBLP:journals/dss/HeinrichZJB21} and deal with small data sets \cite{DBLP:conf/quatic/KappelJS21} ($\sim$) as well as by \cite{DBLP:conf/caise/KlinkmullerBW18} with respect to reliability of the predictions ($\sim$). 

\section{Open Challenges and Research Directions}\label{sec:research_agenda}

The assessment conducted in Sect. \ref{sec:prediction_requirements} shows that some of the CMFs are not supported by existing approaches yet or require a combination of existing PPM approaches. As outlined in the introduction, PCM is a complex and continuous task, requiring a comprehensive support by a PCM system. A comprehensive PCM system requires the realization of predicate prediction and PCM (cf. Fig. \ref{fig:overview_pcm_system}) and the support of the full set of CMFs for both cases.

Overall, we understand PCM as a system that helps a company to understand its future compliance status which encompasses and supersedes all existing approaches.
Understanding PCM as a system results in approaches claiming to develop or implement PCM that these approaches must cover all the extended CMF functionalities and, ideally, also come with a real-world case study that is not simulated using event logs demonstrating the system in practice. Existing event logs such as logs from the Business Process Intelligence Challenge recently organized by the IEEE Task Force on Process Mining\footnote{https://www.tf-pm.org/competitions-awards/bpi-challenge} do not challenge the PCM system in many of the required functionalities, e.g. BPIC logs are already extracted from the information systems. To implement a PCM system, no or almost no abstractions or simplifying assumptions should be made about reality, as these assumptions may decrease the applicability and value in practice. Hence, existing PPM and CM approaches are methods for PCM and only cover fractions of the required functionality (as depicted in Tab. \ref{tab:extended_cmfs1-3} to Tab. \ref{tab:extended_cmfs12-15}). It remains yet to be shown, how the existing approaches can be combined and applied in a case study. If a method is developed that falls into PPM or CM and can be used for PCM, we advocate to phrase it as such that the method helps to realize a PCM system, but in itself is neither PCM nor a PCM system.

Section \ref{sec:prediction_requirements} shows the potential of existing PPM approaches for a comprehensive PCM support. This section summarizes the open PCM challenges and set out research directions for PCM and PPM along the four classes of the extended CMF framework.

\subsection{Modeling requirements}
\label{sub:modeling_sec4}

\noindent\textbf{Holistic modeling and prediction} \\
    \textsl{Challenge:} The modeling and prediction of time, resources and data is supported by existing approaches on predicate prediction and can be  realized by PCM through a combination of PPM with different prediction goals.  What is still missing is the modeling and prediction of case data.  \\ 
     \textsl{Research direction:} 
Modeling requirements are already well-supported. This support can be rounded off with approaches to predict case data, in particular, in the case of IoT data associated with a case \cite{DBLP:journals/corr/abs-2206-11392}. This links to data requirements on the exploitation of context data. Altogether, future work should focus on complex examples from reality (cf. Ex. \ref{ex:eu}) and try to avoid introducing simplifying assumptions or abstractions as much as possible, as these might limit the research to only a subset of modeling requirement CMF combinations. In particular, research should avoid assuming the constraints to be readily available and stated in some logic, but often have to be extracted and updated based on regulatory documents \cite{DBLP:conf/caise/WinterR19}.

\subsection{Execution requirements}
\label{sub:execution_sec4}

\noindent\textbf{Life cycle handling} \\
    \textsl{Challenge:} None of the existing PPM approaches exploits the life cycle of activities, i.e., exploits the semantics of distinct life cycle states/transitions of activities in the event stream. However, these life cycle states/transitions might contribute to predict, for example, the activity duration or might indicate exceptional behavior, resulting in unseen behavior or drift, and subsequently necessitating adequate mitigation actions. \\
    \textsl{Research direction:} Incorporating and exploiting life cycle states and transitions into PPM and PCM might tremendously increase prediction quality and applicability in real-world settings. Consider, for example,
     the transportation of goods to different locations. By distinguishing  the \texttt{start} and \texttt{complete} events of activities, activity duration can be considered in the prediction. By exploiting more life cycle states such as \texttt{suspend} and \texttt{abort}, upcoming exceptions might be predicted and exception handling actions be defined and taken. Assume that, for example, for activity `transport' a \texttt{suspend} event occurs. This might result in delay which should be incorporated in a temporal prediction, e.g., of the remaining time of the affected instance. If an \texttt{abort} event occurs for `transport', we can conclude that the transport will not be completed (i.e., event \texttt{complete} will not occur for 'transport) and this might
result in a compliance violation.
\noindent\textbf{Instance and process spanning constraints} \\
     \textsl{Challenge:} Predicting the compliance of constraints that span multiple process instances and/or multiple processes has not been explicitly addressed by existing approaches, except for encoding inter-case features for improving the prediction. The latter still focuses on predictions of individual instances and not on predictions across multiple instances. 
     However, many application domains crave for compliance support in instance and process spanning settings, e.g., logistics, medicine, and manufacturing. \\
     \textsl{Research direction:} State-of-the-art PPM approaches mostly focus on predicting next activity/event within the context of single instances. However, when considering compliance constraints that span across multiple processes and process instances, it becomes necessary to predict interactions between the affected process instances and their behavior, as well. In particular, such compliance constraints refer to data and/or resources shared by processes/instances. Take as an example the compliance constraint: \textsl{``Each clerk is allowed to issue approve loan as long as a threshold (around \$1M) is not reached. Otherwise he has to delay this event to the following day''} \cite{DBLP:conf/edoc/WinterR17}. First of all, we can see that the constraint imposes a condition across several process instances that refers to data element `threshold' and implicitly to time (`within one day'). Hence, compliance predictions across multiple processes and process instances have to consider a combination of the control flow, data, time, and resource perspective and support the decision on mitigation actions spanning multiple instances and processes. In the example a prediction on the loan amount per clerk across all instances, considering remaining time, can be exploited to trigger proactive ``task switching'' between the clerks, i.e., a clerk with a higher amount of ``free loan'' on this day can swap his loan request with a clerk with an already restricted ``free loan'' amount and a higher loan request. Incorporating the effects of this behavior imposed by the constraints into the prediction is of utmost importance.

\subsection{User requirements}
\label{sub:user_sec4}

\noindent\textbf{Provision of mitigation actions} \\
     \textsl{Challenge:} Though recommendations are used to support users in taking counteractions regarding delays or other risks, none of the approaches suggests  mitigation actions to overcome compliance violations. In particular, approaches are missing that provide mitigation actions at different granularity levels, analyze and visualize the effects of applying mitigation actions, and provide users with estimations on their significance. \\
     \textsl{Research direction:} Based on compliance violation predictions combined with root cause analysis, the effects of mitigation actions can be assessed; either by simulating what will happen if a user applies a specific countermeasure or by determining and suggesting mitigation actions for avoiding the compliance violation. Consider again the transportation example provided in research direction \textsl{life cycle handling}. Assume that based on data gathered before and during transportation, the transportation is predicted to be aborted for a certain process instance. Based on the prediction, we can immediately start to define countermeasures for avoiding the compliance violation of not arriving at the destination, together with predicting their effects. One possible mitigation action in this case is to start another transportation process arriving on time. The prediction can then estimate whether or not the application of this countermeasure compensates the failure. By taking the results of the estimation and an ontology or a knowledge graph 
     either on the domain, i.e., in this case logistics, or on the company executing the process (cf. research direction on explainable, supervised machine learning \cite{burkart_survey_2021}), future work may come up with novel ways to explain the mitigation actions and their effects in an understandable natural language or interpret them as explainable events (cf. event management \cite{chen_event_2020}).
     
\noindent\textbf{Visualization and explanation of predictions and violations} \\
     \textsl{Challenge:} Explainability of prediction results has gained attention. However, visualization approaches for prediction results, especially future compliance violations are mostly missing. Moreover, root cause analysis has to be extended in order to deal with predicting violations of real-world compliance constraints. The challenge becomes event more difficult, when the approach does not stop at some metric quantifying the individual impact of features on the result and visualizing results, but start to tell a story beginning at the root cause and ending at the compliance violation, as coherent, logical storytelling is the means of explaining for humans \cite{burkart_survey_2021}. \\
     \textsl{Research direction:} PCM requires an aggregated view on several perspectives, including the compliance constraints and the process respectively process instance perspective, i.e., a view on the current event stream combined with continuous updates. Though some approaches already provide visualizations for simple SLA violations including color coding, e.g., red means the SLA is violated, green the SLA is not violated, there is still room for improvement and extensions when considering complex compliance constraints. Therefore, visualization approaches are required to depict possible complex information at once, i.e., all compliance states for all processes and process instances captured by the event stream, as well as the definition and visualization of single views, e.g., visualization of compliance predictions for one constraint, one particular instance, or one perspective such as time.  Moreover, information on root causes for compliance violations, the current prediction model in use, and mitigation actions together with their effects should be conveyed to users based on visualization approaches. Through visualization, the PCM system can start to connect trained classifiers with relevant constraints that facilitates fine-grained root cause analysis. 
     At last, the PCM system helps the user in understanding the visualized results by supporting natural language explanations that connect the results in a logical way.
     
\noindent\textbf{Update:} To implement a PCM system, various functionalities for updating parts of the system have to be developed.

\noindent\textbf{Update 1: Treatment of unseen behavior and update}\\
        \textsl{Challenge:} Activity occurrence, absence, and ordering of activities in compliance constraints is covered by existing PPM approaches for those activities that have been already observed. Unseen behavior remains largely uncovered. Unseen behavior can occur in event streams if the underlying process model is not or only partly known or due to concept drift. In combination with compliance constraints, the requirement to predict unseen process behavior becomes even more likely as the compliance constraints do not have to be part of either an underlying process model or the observed behavior in the event stream. Similar observations also hold for the treatment of unseen data and unseen data values (internal and external data). \\
        \textsl{Research direction:} Recently, strategies on how to update the prediction model in case of unseen process behavior in the context of next activity prediction have been proposed, including ``do nothing'', ``retrain without hyperparameter optimization'', ``full retrain'', and ``incremental update'' \cite{DBLP:conf/bpm/PauwelsC21,DBLP:journals/corr/abs-2109-03501,Mangat2022,DBLP:conf/IEEEscc/MaisenbacherW17}. In addition, we need to consider not just updating the model whenever unseen behavior has occurred, but also how to predict the unseen behavior as such, e.g., by considering available context data (cf., e.g., \cite{DBLP:conf/bpm/Francescomarino17}). Therefore, we argue that approaches are not insufficient w.r.t updating models. Dealing with unseen behavior also can mean to predict unseen behavior as it is already done in ML, e.g., zero-shot learning (cf., e.g., \cite{DBLP:journals/pami/XianLSA19} for an overview).\\
       \textbf{Update 2: Online PCM, online (re-)training of prediction models} \\
       \textsl{Challenge:} One situation is to train the prediction model based on available historical information and then constantly update it whenever new information in terms of events or constraints occurs. The more difficult situation is to start from scratch and having to learn and train without any previous knowledge. Both cases of learning/updating the prediction model are challenges for PCM. When considering context data, the decision on when it is necessary to update may be different. \\
       \textsl{Research direction:} As both cases have been addressed by online process mining approaches with respect to new events, research on PCM has to investigate the challenges based on existing work in online process mining. However, approaches to consider new constraints or constraint evolution are missing. Moreover, strategies for updating prediction models at the presence of continuous context data have to be elaborated, e.g., constant updates versus updates if significant changes in the context data occurs.\\
        \textbf{Update 3: Continuous update of prediction results and compliance violations}\label{cha:benchmark} \\
        \textsl{Challenge:} The prediction results and compliance violations have to be updated continuously as new arriving events reveal information about the actual progress of the monitored process(es). Here, the challenge is to investigate whether the performance of the PCM system allows to update all results for each new incoming event or whether batching of incoming events is necessary. These considerations have to take the (end) user requirement on the performance of the approach into account. The challenge becomes more difficult, if the evolution of the constraints base through, for example, changing regulatory documents (cf. Ex. \ref{ex:eu}) is also taken into account.  \\ 
        \textsl{Research direction:} 
        This challenge requires thorough case studies that result in a comprehensive benchmarking environment for PCM systems. The case studies should come with time series data on requirements (simulating the evolution of requirements during development and operations of a PCM system), regulatory documents, relevant and irrelevant context and the raw events from various information systems, possibly also from various processes even spanning multiple partners. Only then can proposed PCM systems be properly compared.  \\
        \textbf{Update 4: Compliance degree and update} \\
        \textsl{Challenge:} First approaches for predicting the compliance degree across multiple process instances have been presented. Yet, especially in combination with updating compliance violations, an open challenge remains how to define and update the compliance degree while new events arrive throughout the event stream and to predict compliance states of single instances. Furthermore, it remains unclear how these instance-level compliance states and degrees can be lifted to the process or choreography level. As prediction results include probabilities, e.g., how likely is the occurrence of a certain activity, these probabilities should be considered for compliance states and degrees. \\
        \textsl{Research direction:} Following \cite{maggi_compliance_2019}, currently we distinguish compliance states \textsl{possibly violated/satisfied} and \textsl{violated/satisfied} for single instances. \textsl{Possibly violated} means that the violation can be still healed, i.e., by the occurrence of an activity that is mandatory according to the compliance constraint (cf. \cite{DBLP:journals/isf/LyRGD12}). A (final) violation, by contrast, states that the constraint cannot be healed anymore, e.g., if the constraint is \textsl{possibly violated} and then the end event of a process instance or all end events occur. There is also a distinction between \textsl{fully} and \textsl{partly violated/satisfied} where full violation of a compliance constraint means that this constraint is violated for all process instances, and a partly violation means that the constraint is violated for at least one constraint \cite{DBLP:conf/bpm/TosattoGB19}. It is unclear whether the proposed compliance states are actually sufficient in light of the prediction and continuous update performed by PCM systems. In particular, the presence of probabilities on prediction and, thus, compliance state results expressing the reliability and certainty of the PCM system on the results may necessitate further compliance states that communicate relevant aspects of the results to the end user in a compact and understandable way. For more complex compliance constraints referring to activities, data, time, and resources, the probabilities of satisfying/violating these constraints have to be calculated in an adequate manner, e.g., how likely is the occurrence of a certain activity producing a certain data value in a given time span? If these probabilities can be determined, in turn, the risk of violations can be assessed and the compliance state properly identified. Compliance states may also need to express boundary considerations (based on the information we have, what has happened in the best and worst case at the partner?) when considering private processes in process choreographies. In connection with the challenges on data quality, a compliance state may need to also reflect the uncertainty on the process or context data, as data sources such as sensors can carry information on their accuracy. For uncertain process data, PCM system research should investigate based on recent work on data quality in the data preparation phase \cite{wynn_responsible_2019} or in conformance checking \cite{felli_conformance_2022}. By defining a suitable aggregation mechanism, work on compliance states and degrees can be lifted to the process or cheoreography level. Somewhat orthogonal to the aforementioned lines of research is the question on how to update the compliance degrees and states as prior parts of the PCM systems (see above) are updated. This remains unsolved so far.

\subsection{Data requirements}
\label{sub:data_sec4}

    \noindent\textbf{Heterogeneous data from distributed processes and (contextual) data sources}\\
     \textsl{Challenge:}
     First approaches start addressing data requirements in PPM and PCM. However, approaches for event and constraint data from distributed and heterogeneous sources and processes are missing. Moreover, the ongoing exploitation of contextual data is promising, but under-researched yet. \\
    \textsl{Research direction:} Future research on heterogeneous data for PCM may have to borrow techniques from data integration research that deal with the problems of, for example, entity recognition, data fusion or schema alignment \cite{dong_data_2018} before the line of research that tackles the problems of event extraction, event abstraction and handling event data that does not come with a single case identifier can be followed. In particular for event data that does not come with a single case identifier, a new field of research within process mining called object-centric process mining that conceptualizes such event data as object-centric event logs (OCEL) \cite{ghahfarokhi2021ocel} is emerging. These problems are likely to occur, if relevant data on processes is distributed across multiple information systems either within a company or among partners. These problems are aggravated when compliance constraints are not stated in a machine-readable format carrying semantics of a logic such as LTL. Typically, constraints are stated in natural language and scattered across multiple regulatory documents (cf. Ex. \ref{ex:eu}). Then, the regulatory documents are further data sources that are challenging to include, as constraints have to be extracted in a way that we can match events/activities with them. Including context data potentially coming from other sources than the process data into PCM can significantly increase prediction capabilities and quality. In manufacturing processes, for example, several sensor data streams are measured continuously that report the environment state/context of the process, e.g., the temperature of a room/machine or the fluid level in the machine. Detecting deviations in the context data can increase prediction effectiveness, e.g., concept drifts might be predicted early \cite{DBLP:conf/bpm/StertzRM20}. Additionally, data value prediction might not only refer to process data, but also to contextual data such as time series, for example, if decision rules are based on time series data such as temperature \cite{scheibel2022timeseries}. Here, the combination of PCM with time series prediction approaches constitutes a promising research direction \cite{DBLP:journals/cm/HuaZLCLZ19}. \\
     \noindent\textbf{Prediction and compliance in process choreographies} \\
     \textsl{Challenge:} This challenge is related to the previous challenge on data from multiple sources as the choreography of partner processes results in heterogeneous data sources, e.g., one data source per partner. Especially for choreographies, additional challenges arise from confidentiality requirements of the partners, i.e., aggravates the challenges for multiple, heterogeneous data sources by restricted visibility of the data. As a consequence, as there can exist constraints spanning multiple partners, predicting compliance in this distributed setting is a task that typically cannot be performed at one partner’s side. Hence, process choreographies pose new challenges for PCM such as how the prediction and checking of constraints is actually performed (in a distributed way?). If private processes of partners come into play, the prediction of and checking of constraints has to adhere to privacy and confidentiality requirements. \\
     \textsl{Research direction:} Although first approaches enable to check compliance in process choreographies (e.g., \cite{fdhila2022verifying}), the problem of predicting compliance in process choreographies has not been addressed so far. One challenge arises from the confidentiality requirements of the private partner processes: if they are affected by compliance predictions, it might become necessary to distribute the prediction  among the partners. This endeavour may be realized by means of \textsl{secure multi-party computation} \cite{lindell_secure_2021}. \\
     \noindent\textbf{Data properties and quality} \\
     \textsl{Challenge:} The exploitation of data properties and quality is promising. One drawback of existing PPM approaches with respect to the input data is the assumption of label equivalence, i.e., the prediction are based on labels of events. Label equivalence is not sufficient, particularly when merging event streams from heterogeneous input sources (variety). Another challenge is the size of the input data which can be too small or too big (volume). For distributed processes, event streams might contain information on message exchanges between partners --how can they be exploited for prediction or being predicted themselves?-- and might also contain hidden/invisible parts due to confidentiality requirements of the partners. Here, an initial hurdle is the lack of data sets. Another one is that event logs in practice are error-prone \cite{DBLP:journals/is/CappielloCPF22}. When considering data quality, the data values of low quality may signify a compliance violation or be used as a feature for prediction. Determining whether the former is the case and the latter is beneficial, is yet another challenge. \\
     \textsl{Research direction:} To tackle variety challenges, PPM and PCM can benefit from equivalence notions that aim at the functionality of activities, e.g., attribute equivalence \cite{koenig_compliance_2019}. First approaches for dealing with volume challenges boost small data sets \cite{DBLP:journals/infsof/KappelSJ21,DBLP:conf/caise/HerbertMR21}. Other approaches aiming at efficiency and performance of PPM and PCM with respect to both, volume and high velocity event streams are missing (note that for PCM also a large set of compliance constraints might exist). To deal with PCM in distributed settings, the collection, provision, and preparation of (real-world) data for different process scenarios containing multiple perspectives remains an ongoing direction of research. With respect to data quality and its exploitation in a PCM system, approaches are missing. Here, PCM can investigate based on work considering data quality as uncertain data values \cite{wynn_responsible_2019,felli_conformance_2022}.  \\

\subsection{Overall}

    \noindent\textbf{Systematic assessment of data mining/machine learning techniques}\label{cha:ml} \\
    \textsl{Challenge:} In the light of a multitude of challenges and research directions and the lack of a single prediction technique implementing or supporting all extended CMFs, it remains a challenge to apply an existing or develop a new prediction technique that addresses all of the challenges and implements or supports all extended CMFs. Due to the many facets of the problem and functionalities required, it is likely that not each problem can be tackled by existing data mining/machine learning techniques. \\
   \textsl{Research direction:} First, a systematic assessment of (existing) prediction techniques to address these challenges is required. As it is likely, that not all of the challenges for PCM can be solved by existing prediction techniques, it might be necessary to develop new techniques. Another possible direction could be to decompose the problem and reformulate it either as a \textsl{multi-view learning} problem or a \textsl{multi-task learning} problem \cite{zhang_survey_2021}.

\noindent\textbf{Predicate Prediction vs. PCM}\label{cha:predvspcm} \\
    \textsl{Challenge:} Throughout the paper, we have highlighted two options that can be used in a PCM system, i.e., \textsl{predicate prediction} or \textsl{PCM}, cf., Fig. \ref{fig:overview_pcm_system}. However, it remains unclear when to use which of those options, in particular, as there remain several challenges and open research gaps for the latter. \\
   \textsl{Research direction:} After having achieved a clear understanding of PCM, we can carry out several comparisons between both options. First, they can be compared by measuring, e.g., accuracy of both in the same setting of constraints. One could imagine that when one constraint never changes it might be preferable to use predicate prediction in the sense that higher accuracy could be achieved. However, this needs to be investigated in detail, through, e.g., cases studies which could result in recommendations which method to prefer in which setting.

\section{Discussion and Conclusion}\label{sec:discussionandconclusion}

\subsection{Discussion}\label{sec:discussion}

The survey aims at shedding light on the current position of existing CM and PPM approaches with respect to the extended CMF framework. Moreover, based on an assessment of existing CM and PPM approaches, it addresses the question whether their combination solves the challenge of generating a PCM system in its entirety. We followed selected principles of conducting a systematic literature review \cite{DBLP:journals/jss/BreretonKBTK07} and adapted them in terms of incorporating existing surveys as basis whenever possible. In this spirit, we took the established CMF framework \cite{DBLP:journals/is/LyMMRA15} and extended it based on more recent findings. Despite this careful method design, the following limitations can be identified.

\begin{itemize}
\item New approaches on PPM are published constantly. Therefore, one limitation of this work is that new approaches since the literature compilation in February might have been published which are consequently not covered within this paper yet. A search on Google Scholar with \texttt{allintitle:predictive process monitoring} and selecting papers after $2022$ results in $31$ hits\footnote{accessed 2022-10-20}. From these $31$ hits, this survey covers \cite{rizzi_explainability_2020,DBLP:journals/kbs/JalayerKPB22,DBLP:journals/dss/KimCDMT22}. $28$ papers are not covered, out of which $2$ are out of scope following our search methodology in Fig. \ref{fig:lit_comp_wf}, $15$ papers have been published as technical reports and $1$ as PhD thesis. Looking a this most recent work, the majority of the approaches is concerned with explainability, some combined with data issues such as \cite{DBLP:journals/corr/abs-2202-08041} (cf. CMF9, CMF14, and CMF15) and with updating the prediction model (cf. CMF8).
\item  The main focus of this paper is on prediction tasks and compliance monitoring. Hence, further related areas such as online process mining, concept drift detection, and anomaly detection approaches have only been considered if papers from these areas were detected during the systematic literature review. Online process mining is often geared towards concept drift detection. This work covers several concept drift detection approaches \cite{DBLP:conf/IEEEscc/MaisenbacherW17,DBLP:conf/bpm/StertzRM20}, also in connection with updating prediction models at the presence of concept drift \cite{DBLP:journals/corr/abs-2109-03501,DBLP:conf/bpm/PauwelsC21}. Anomaly detection can provide insights to PCM. However, most anomaly detection approaches work offline, some can be applied on event streams, e.g., \cite{DBLP:journals/is/KoC22}, but process anomaly predictions beyond the approaches studied in this work such as \cite{DBLP:journals/is/BohmerR20} are missing.
\item  If surveys exist for the investigated research areas, i.e., for CM and PPM, we used these surveys as a basis for our further literature analysis. Doing so might result in missing papers that have been published prior to the existing surveys and not having be treated by them. However, we conducted a full search without restricting the publication dates first and then compared the identified set of papers to existing surveys. Doing so, we limit the risk of missing out relevant prior work.
\item The aim of this paper is not to assess PPM approaches in terms of machine learning or data mining techniques in detail, i.e., the goal is not to identify the PPM approach currently performing best. Instead, the goal is to provide a comprehensive outline and analysis of the PCM system problem and how it is addressed by the current literature. Hence, at this point, we do not investigate or propose particular techniques from a technical point of view. This endeavour is left as a Challenge \ref{cha:ml} to future work.
\item   Though there are case studies for compliance monitoring available ($\mapsto$ Sect. \ref{subsec:findings_casestudies}), we still need to have a detailed look and investigate whether these are suitable for evaluating approaches tackling the mentioned research directions, i.e. they meet the requirements set in Challenge \ref{cha:benchmark} for a benchmark case study.

\end{itemize}

\subsection{Conclusion}\label{sec:conclusion}

This work provides a comprehensive analysis of existing CM and PPM research in in terms of the extended CMF framework and the realization of a comprehensive PCM system. We tackled research questions RQ1 -- RQ4 (cf. Sect. \ref{sec:introduction}) as summarized in the following. In addition to findings on the relationship between CM, PPM and PCM, the study particularly provides findings on PPM and its capabilities. \\

\noindent\textsl{RQ1: To what extent is a PCM system in its entirety addressed and solved by existing PPM and CM approaches?}
RQ1 is addressed by an extensive compilation of literature on PPM, CM, and PCM. Based on analyzing the literature, we conclude that the PCM system in general has not been developed by now. In particular, the integration of predicate prediction and PCM in one system is missing, including an understanding of their pros and cons. \\

\noindent\textsl{RQ2: How are the existing PPM and CM approaches comparable in terms of CMF functionalities, in particular, in terms of PCM system requirements necessary for the respective functionality?}
The selected literature from PPM, CM and PCM research emphasizes that the compliance monitoring functionalities as originally proposed by \cite{DBLP:journals/is/LyMMRA15} in 2015 are still valid and can serve as requirements for the PCM system. The CMF framework is extended based on analyzing the literature compilation regarding PPM and CM directions after 2015, most prominently, towards the explainability of the prediction results and input data requirements. To further the predictive capabilities necessary for PCM, the extended CMF framework is analyzed for predictive requirements arising for each of the CMFs, illustrated by means of Ex. \ref{ex:eu}, and is subsequently used to assess existing mostly PPM approaches. \\  
     
\noindent\textsl{RQ3: Which of the PCM system requirements are already met by existing approaches?}
In general, PPM holds the capabilities to tackle PCM challenges and can be seen as an enabler for PCM system development. These capabilities are derived for each CMF functionality of the extended CMF framework and used to assess existing approaches. The assessment finds capabilities to be (partly) supported, but there is no comprehensive solution for all PCM system requirements, i.e., the assessment eventually results in a list of open PCM system challenges, including the integration of heterogeneous data sources as well as handling distributed processes and more complex constraints and their monitoring. In both, CM and PPM research, the system development aspect is missing to a large extent. \\
    
\noindent\textsl{RQ4: Which open challenges and research directions remain for full PCM support?}
Based on the identified open PCM challenges, together with the PCM system (prediction) requirement list, research directions for PCM are elaborated. These research directions comprise the holistic prediction of activities with data, time and resources, an appropriate life cycle handling, a support for instance- and process-spanning constraints, the provision of mitigation actions, the explainability of compliance violation predictions, multiple functionalities to update various parts of the PCM system containing a call for a thorough case study necessary to benchmark approaches, the capability to deal with process choreographies and the explicit treatment of data quality and properties, also for heterogeneous processes and data sources. All of these open challenges constitute key success factors for predictive compliance monitoring. 

The research directions point to several future research opportunities. 
Working on the research directions will necessitate a comprehensive assessment of existing machine learning and data mining techniques and might result in the development of extended or even new techniques. Moreover, this work assumes that compliance constraints are formalized using some notion. In future work, we will incorporate the sources, e.g., regulatory documents, into predictive compliance monitoring. In order to evaluate and compare new techniques and approaches, appropriate data sets are crucial, i.e., event streams, contextual data, data from different sources and processes, and unseen data.

\section*{Acknowledgments}
This work has been supported by Deutsche Forschungsgemeinschaft (DFG), GRK 2201 and by the Austrian Research Promotion Agency (FFG) via the Austrian Competence Center for Digital Production (CDP) under the contract number 881843.


\appendix

\newpage

\section{Appendix}
\label{sec:appendix}
\begin{figure}[ht!]
    \centering
    \includegraphics[width=\textwidth]{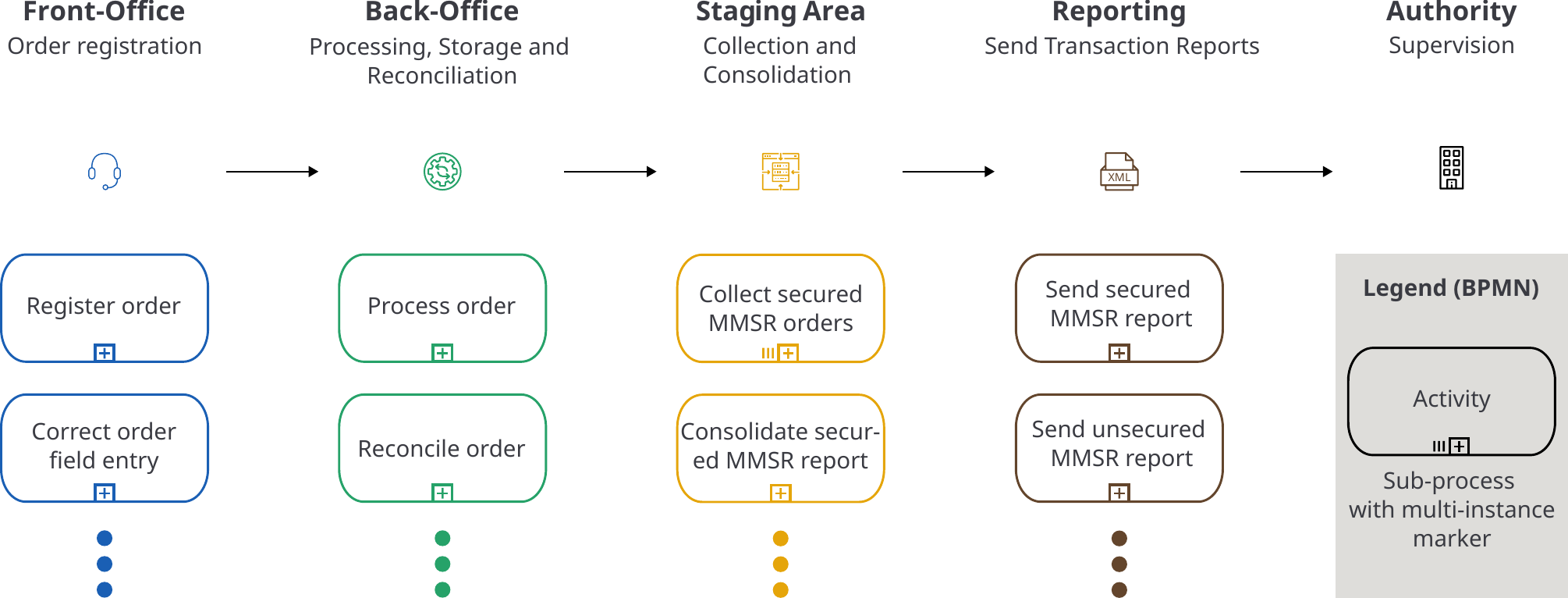}
    \caption{Transaction Reporting Processes of Financial Institutions in the EU \cite{thomas_wenzel_transaction_2020}}
    \label{fig:example_process}
\end{figure}

\begin{figure}[ht!]
    \centering
    \includegraphics[width=\textwidth]{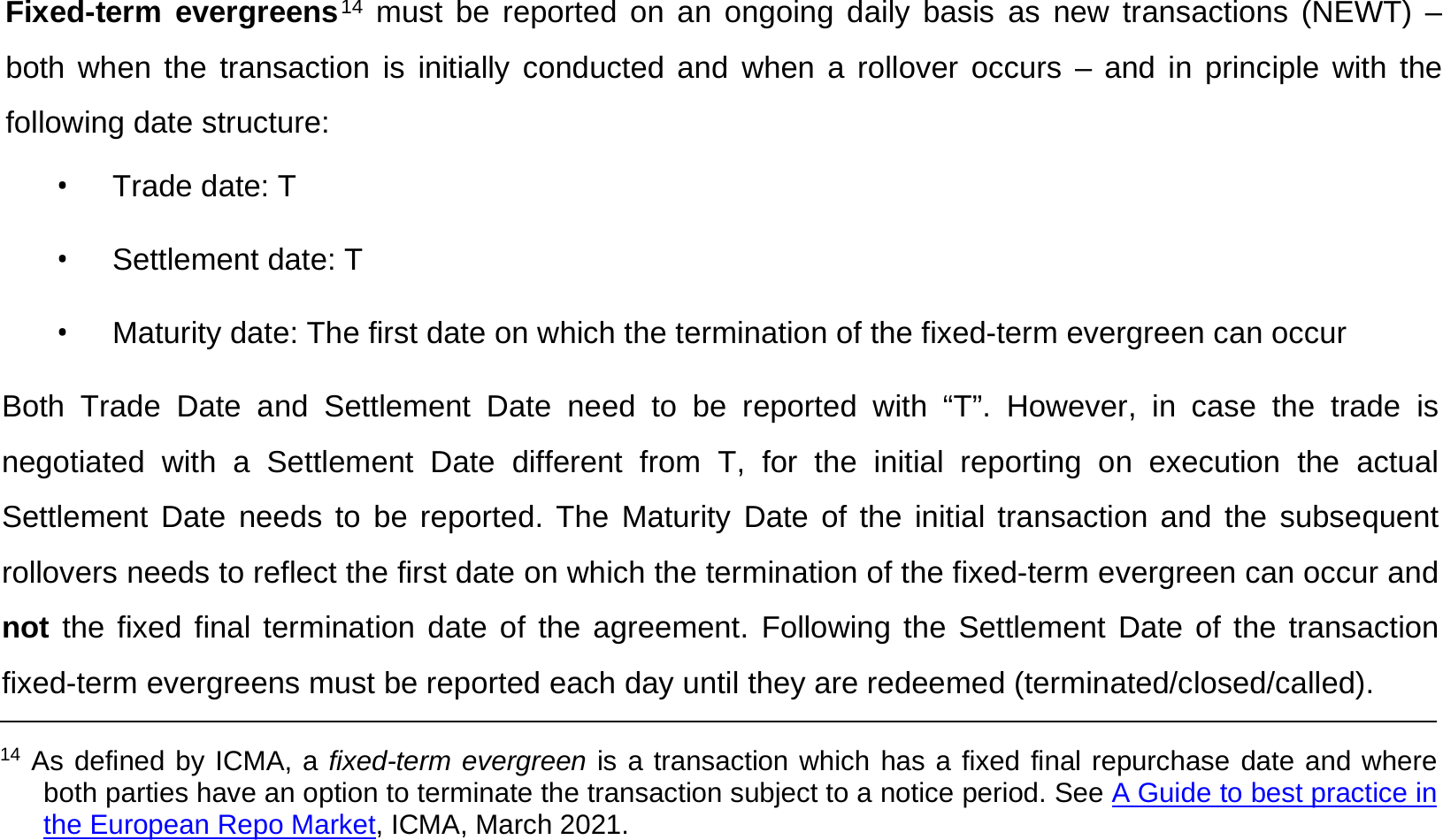}
    \caption{Constraint for Case Data of an Order \cite{european_central_bank_reporting_2021}}
    \label{fig:constraint}
\end{figure}

\begin{example}[Transaction Reporting of Financial Institutions in the EU]\label{ex:eu}
Beginning in 2014, authorities such as the European Commission, the European Parliament, the European Central Bank, the European Banking Authority and the European Securities and Markets Authority imposed regulations on financial institutions conducting money and capital market business in the EU to report money and capital market transactions on a daily basis to the respective authorities as a prerequisite to continue conducting that type of business. The required daily reports are based on the European Markets Infrastructure Regulation (EMIR) \cite{european_parliament_regulation_2014}, Money Markets Statistical Reporting (MMSR) \cite{european_parliament_regulation_2015}, Securities Financing Transactions Regulation (SFTR) \cite{european_parliament_regulation_2019} and Markets in Financial Instruments (MiFIR/D) \cite{european_parliament_regulation_2018}. These regulations are typically complemented by multiple addenda and technical specification documents that further clarify the exact requirements the institution has to fulfill, e.g., for MMSR the reporting instructions \cite{european_central_bank_reporting_2021}, questions and answers \cite{european_central_bank_money_2021}, IT appendix \cite{european_central_bank_mmsr_2017}, data quality checks \cite{european_central_bank_moneymarket_2021} and further technical specifications documents for the web service and XML schema available for download as a ZIP\footnote{\label{footnote1}\url{https://www.ecb.europa.eu/stats/financial_markets_and_interest_rates/money_market/html/index.en.html}}. All in all, the MMSR regulation stipulates the reporting of four different MMSR reports, among others the secured and unsecured market segment reports, that are used to determine the euro short-term rate (\euro{}STR), an important interest rate. \\ The regulatory documents specify the outcome of the process and necessitate the occurrence of certain activities, but do not specify the actual process model in full detail. Hence, the specification gives financial institutions similar to specifications in the healthcare domain \cite{kaes_flexibility_2014,kaes_acaplan_2015} significant flexibility for implementing and executing the process with individual subprocesses for activities \cite{thomas_wenzel_transaction_2020}. Nevertheless, by abstracting the individual subprocesses of institutions, similar activities for the transaction reporting process can be deduced as depicted in Fig. \ref{fig:example_process}. In the following, the coarse-grained activities that are similar to all financial institutions are described.\\Traders or Sales personnel in the front-office agree on a transaction with an external trading party through electronic trading software, e.g., Bloomberg, or over the phone. For each transaction, an order is registered or has to be registered by the trader (``Register order''). If at some point a field of the order has been filled incorrectly, the trader has to correct the entry (``Correct order field entry''). As an agreed transaction needs to be cleared, i.e., payments and the actual transfer of owner/holding rights with respect to the depository need preparation and execution, it is processed in the middle-office (``Process order''). In order to prevent fraud and guarantee clearing according to previously agreed upon terms by both parties of a transaction, orders are reconciled (``Reconcile order''). In order to transform the large number of orders from various electronic order information systems to the specified report structure and format, all secured market segment orders are collected for the secured MMSR report (``Collect secured MMSR orders'') and consolidated (``Consolidate secured MMSR orders''). Afterwards, the final report can be sent to the respective authority (``Send secured MMSR report''). \\ According to Annex IV 2. (i-ii) and 3. (iii) the reports have to be accurate, the reporting process must be monitored, and any deviation needs to be explained by the institution within $45$ min to $7$h depending on the time of occurrence \cite{european_parliament_regulation_2015,european_central_bank_money_2021}. According to Article 4 1. (a), all MMSR reports have to be sent between $6$pm of the same trade day that the transaction was agreed upon and $7$am of the following trade day. In \cite{european_central_bank_mmsr_2017}, the ``Send [type] MMSR report'' activity is specified. In Fig. \ref{fig:constraint}, an extract from \cite{european_central_bank_reporting_2021} shows a constraint that specifies for fixed-term evergreens as defined by the International Capital Markets Association (ICMA) with collateral having a valid International Security Identifcation Number (ISIN), i.e. money market loan contracts for which the borrower pledges another, registered capital market product to secure the repayment of the loan, the reporting in the secured market segment and attribute values ``T'' for both ``Trade date'' and ``Settlement date''. Further exemplary constraints \ref{ex_c1} -- \ref{ex_c16} were introduced in Sect. \ref{sec:prediction_requirements}.

All in all, the regulatory documents for MMSR span hundreds of pages containing a multitude of constraints with varying detail. The financial institution is responsible to transform and map all of these requirements to its internal reporting process that adheres on a high-level to the previously described reporting process. Furthermore, all of the documents can change and have changed since the regulation became in force. For example, the regulation itself was amended four times and the reporting instructions at least six times$^{\ref{footnote1}}$. These changes lead to concept drifts in the reporting process, as historically possible activities may or cannot occur anymore and at the same time new activities might 
have to be executed that have not been observed before; these activities
account for unseen behavior \cite{DBLP:conf/bpm/PauwelsC21,DBLP:journals/corr/abs-2109-03501,Mangat2022}. Additionally, interface changes of information systems for electronic trading that are used internally or by external trading parties (e.g. exchanges), as well as other interrelated IT system changes can lead to concept drift in the reporting process. \\ To sum up, the financial institution has to implement a complex transaction reporting process with a large number of constraints that runs every trade day, has to monitor its execution and explain deviations. \end{example}

As described in Example \ref{ex:eu}, the financial institution has to monitor the compliance of the reporting process to the external regulatory constraints and further internal constraints necessary to comply to the governance guideline by the European Banking Authority \cite{european_banking_authority_guidelines_2018}. To comply to the punctual sending of an accurate report and to proactively manage possible deviations due to IT incidents, wrong data attributes or external factors such as days with exceptionally high order volume, it also needs to ultimately monitor the remaining time prediction to complete the sending of the report and the data attributes predictions and life cycle of that activity with respect to the relevant constraints. In addition, because of concept drifts due to changes in various regulations, various information systems internally and externally, and IT incident and behavior patterns, a PCM system is required that regularly updates both the constraints set and the prediction and is also able to predict unseen behavior, for example data attributes in the report for which the historical data has never recorded a corresponding order. As the financial institution is required to explain deviations, it needs a root-cause analysis functionality and, because of the short period of time for explaining the deviations, ideally some functionality to explain what the results of the system mean.

To further exemplify the transaction reporting process with respect to PPM and CM, in the following, we present two scenarios for Example \ref{ex:eu} that represent two days in which violations of the constraint might occur:

\begin{scenario}[Correct orders]\label{sce:orders} Consider trade day 29.09.2022, on which for the transaction reporting process of the German BANK the activities for processing, storage and archiving for the MMSR reports (Middle-Office and Back-Office) that usually occur from 8pm to 2am are running one hour longer than they used to run and the data collection and consolidation activities run 4 hours on average such that a violation of the MMSR regulation timeliness constraint is likely. Assume that the longer running time for processing, storage and archiving is due to an exceptionally high number of "Correct order field entry" activities. \end{scenario}

\begin{scenario}[Conflicting rules]\label{sce:rules} Consider trade day 30.09.2022 6pm, at which time the reporting department employee on-call duty has already worked 10 hours. Nevertheless, the report processing breaks at 7pm of the same day such that the employee on-call duty has to intervene. Assume the 10 hour day of the employee was due to the assessment of protection requirements according to \cite{bundesanstalt_fur_finanzdienstleistungsaufsicht_bankaufsichtliche_2021} and the breaking of the processing due to a hardware anomaly on one of the processing servers. \end{scenario}

In the first scenario, following the aforementioned requirements for monitoring the reporting process, the BANK wants to know and understand in particular how likely it is that the MMSR report is sent in time, whether this likelihood changes over night, what the root cause for the longer running Middle- and Back-Office activities is and how to mitigate the potential violation of the timeliness constraint in the short- and long-term. In the second scenario, the BANK has to additionally understand, that it either can meet the requirement of the MMSR reporting process by fixing the broken processing or the requirement for German employees not to work more than 10 hours per day (\S 3 of the German ``Arbeitszeitgesetz''). 
\end{document}